\documentclass[useAMS,usenatbib]{mn2e}

\usepackage{verbatim} 
\usepackage{natbib} 
\usepackage{amsmath} 
\usepackage{amsbsy}
\usepackage{amssymb}
\usepackage{mathrsfs} 
\usepackage{lscape} 
\usepackage{graphicx}
\usepackage{epstopdf}
\usepackage{deluxetable}
\usepackage{ctable} 
\usepackage{fixltx2e}

\title[Bursty Star Formation in Galaxies]{A Model for the Origin of Bursty Star Formation in Galaxies}
\author[Faucher-Gigu\`ere]{Claude-Andr\'e Faucher-Gigu\`ere,$^{1}$\thanks{cgiguere@northwestern.edu}
}

\begin{document}
\maketitle

\begin{abstract}
We propose a simple analytic model to understand when star formation is time-steady versus bursty in galaxies. 
Recent models explain the observed Kennicutt-Schmidt relation between star formation rate and gas surface densities in galaxies as resulting from a balance between stellar feedback and gravity. 
We argue that bursty star formation occurs when such an equilibrium cannot be stably sustained, and identify two regimes in which galaxy-scale star formation should be bursty: i) at high redshift ($z\gtrsim1$) for galaxies of all masses, and ii) at low masses (depending on gas fraction) for galaxies at any redshift. 
At high redshift, characteristic galactic dynamical timescales become too short for supernova feedback to effectively respond to gravitational collapse in galactic discs (an effect recently identified for galactic nuclei), whereas in dwarf galaxies star formation occurs in too few bright star-forming regions to effectively average out. 
Burstiness is also enhanced at high redshift owing to elevated gas fractions in the early Universe. 
Our model can thus explain the bursty star formation rates predicted in these regimes by recent high-resolution galaxy formation simulations, as well as the bursty star formation histories observationally-inferred in both local dwarf and high-redshift galaxies. 
In our model, bursty star formation is associated with particularly strong spatio-temporal clustering of supernovae. 
Such clustering can promote the formation of galactic winds and our model may thus also explain the much higher wind mass loading factors inferred in high-redshift massive galaxies relative to their $z\sim0$ counterparts. 
\end{abstract}

\begin{keywords}
Galaxies: formation, ISM, starburst, high-redshift, dwarf -- stars: formation
\vspace{-0.5cm}
\end{keywords}

\section{Introduction}
\label{sec:intro}
Observations of relatively tight relationships between the star formation rates (SFR) of galaxies and their stellar masses \citep[$M_{\star}$; e.g.,][]{2007ApJ...660L..43N, 2012ApJ...754...25R, 2014MNRAS.443...19R} 
and between their star formation rates and gas surface densities \citep[the Kennicutt-Schmidt (KS) relation; e.g.,][]{1998ApJ...498..541K, 2010MNRAS.407.2091G} indicate that star formation must proceed relatively smoothly in typical galaxies when the SFR is averaged over long timescales ($\gtrsim 100$ Myr). 
In most galaxy formation models to date, including large-volume cosmological hydrodynamic simulations \citep[e.g.,][]{2003MNRAS.339..312S, 2014MNRAS.444.1518V, 2015MNRAS.446..521S, 2016MNRAS.462.3265D}, semi-analytic models \citep[e.g.,][]{2013MNRAS.434.1531F, 2014MNRAS.444.2599B, 2015MNRAS.453.4337S, 2015MNRAS.451.2663H}, and analytic ``equilibrium'' models tied to gas accretion rates from the intergalactic medium \citep[e.g.,][]{2010ApJ...718.1001B, 2012MNRAS.421...98D, 2013ApJ...772..119L}, star formation in individual galaxies is in fact effectively assumed to proceed smoothly in intervals between disturbances like galaxy mergers. 

\begin{figure}
\begin{center}
\includegraphics[width=0.475\textwidth]{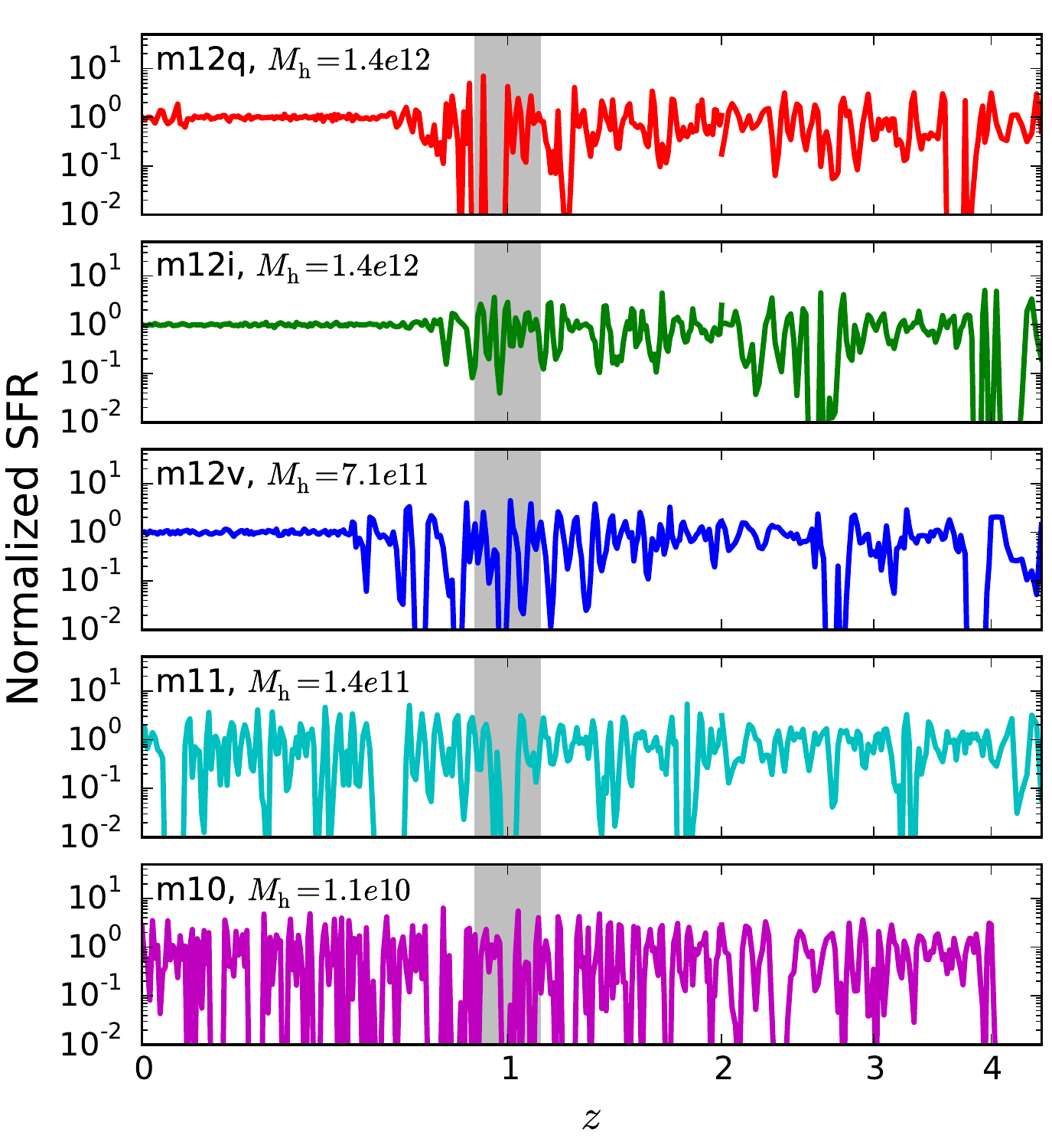}
\end{center}
\caption[]{Normalized SFR versus redshift for simulated galaxies from the FIRE project, in decreasing order of $z=0$ halo mass (labeled at the top left of each panel) from top to bottom. 
The instantaneous SFR is normalized by the running mean SFR, boxcar averaged over $\approx 300$ Myr. 
At high redshift ($z \gtrsim 1$), all simulated galaxies exhibit bursty star formation. 
The more massive galaxies settle into a more time-steady mode of star formation at lower redshifts but the dwarf galaxies sustain bursty star formation down to $z=0$.  
Data from \citet{2015MNRAS.454.2691M}.
}
\label{fig:FIRE_SFHs_norm} 
\end{figure}

However, several recent galaxy simulations with resolution sufficient to resolve the gravitational collapse of individual gravitationally-bound clouds (GBCs) in the interstellar medium (ISM) and to model stellar feedback on the scale of individual star-forming regions predict much more variable star formation histories in some regimes. 
Bursty star formation\footnote{In this paper, ``bursty'' refers to galaxies in which a significant fraction of star formation occurs in recurrent bursts, even if the SFR is time steady when averaged over cosmological timescales. This is in contrast, for example, to isolated bursts of star formation triggered by galaxy mergers.} at high redshift and in dwarf galaxies is in particular a key prediction of the FIRE cosmological zoom-in simulations, which model stellar feedback in a spatially and temporally resolved manner \citep[e.g.,][]{2014MNRAS.445..581H, 2015MNRAS.449..987F, 2015MNRAS.454.2691M, 2017MNRAS.466...88S}. 
This is illustrated in Figure \ref{fig:FIRE_SFHs_norm}, which shows star formation histories for five simulated galaxies from the FIRE project, ranging from dwarf galaxies to Milky Way-mass galaxies, normalized by the running mean SFR averaged over a timescale $\approx 300$ Myr. 
All FIRE galaxies are bursty at high redshift ($z\gtrsim1$, though with significant dispersion in the transition redshift). The more massive simulated galaxies (halo mass $M_{\rm h} \sim 10^{12}$ M$_{\odot}$) settle into a more time-steady mode of star formation at $z\lesssim 1$, while lower-mass galaxies continue to be bursty all the way to the present time. 
In simulations of $\sim L^{\star}$ galaxies, the transition from bursty star formation at high redshift to more time-steady star formation at later times seems to be associated with the transition from highly dynamic and morphologically disturbed galaxies to more well-ordered discs familiar from observations of the nearby Universe \citep[e.g.,][]{2014MNRAS.445..581H, 2015ApJ...804...18A}.

Our goal in this paper is to develop an analytic model to understand when and where galactic star formation is expected to be bursty vs. time steady. We seek in particular to explain the results of simulations like the FIRE simulations. 
We however stress that similar bursty star formation is not limited to the FIRE simulations but is also seen in many other high-resolution simulations using different codes \citep[e.g.,][]{2012MNRAS.422.1231G, 2013MNRAS.429.3068T, 2015ApJ...804...18A, 2015MNRAS.451..839D}.

Bursty star formation has particularly important implications for dwarf galaxies, as simulations indicate that the accompanying time-variable outflows can transfer energy to the central parts of dark matter halos. This process produces cored dark matter halo profiles, resolving a primary tension between the predictions of pure cold dark matter simulations and observations of dwarf galaxies \citep[e.g.,][]{2012MNRAS.421.3464P, 2014ApJ...789L..17M, 2015MNRAS.454.2092O, 2015MNRAS.454.2981C, 2016ApJ...827L..23W, 2016arXiv161102281F}. 
Bursty star formation continuing to late times in dwarfs causes their outflows to recycle many times \citep[][]{2017MNRAS.470.4698A}, which likely plays an important role in maintaining a population of dwarf galaxies blue to $z\sim0$, a challenge in many galaxy formation models \citep[e.g.,][]{2015MNRAS.451.2663H}. 

The burstiness of star formation identified in recent galaxy formation simulations may explain several observational indications of time variable star formation on different timescales. 
Observationally, the burstiness of star formation can be probed by comparing SFR measurements using indicators sensitive to different timescales \citep[][]{2012ApJ...744...44W, 2015MNRAS.451..839D, 2017MNRAS.466...88S}. 
Two of the most used indicators for this purpose are the H$\alpha$ nebular optical recombination line and the ultraviolet (UV) continuum. 
Whereas H$\alpha$ is excited by ionizing radiation from the most massive stars and is sensitive to SFR variations on timescales $\lesssim 5$ Myr, the UV continuum is due to non-ionizing photospheric emission from stars with lifetimes up to $\sim 300$ Myr. 

In the local Universe, 
observations show that the scatter in the H$\alpha$-to-UV ratio increases with decreasing galaxy mass \citep[e.g.,][]{2012ApJ...744...44W}. 
By modeling the observations using toy star formation histories, \cite{2012ApJ...744...44W} showed that the increased 
scatter toward low masses can be explained by increasing burstiness in dwarf galaxies.\footnote{The mean H$\alpha$-to-UV ratio decreases with decreasing galaxy mass, which can also be interpreted as a signature of bursty star formation: if time intervals between bursts are sufficiently long, H$\alpha$ will be depressed relative to the UV continuum most of the time. 
} 
Using a different observational approach combining the 4000~\AA~break and H$\delta_{\rm A}$ stellar absorption line indices with SFR/$M_{\star}$ derived from emission line measurements, \cite{2014MNRAS.441.2717K} also found that the burstiness of recent star formation increases from $M_{\star} \sim 10^{10}$ M$_{\odot}$ to $\sim 10^{8}$ M$_{\odot}$. 
Interestingly, this technique probes variability on longer times scales, $\gtrsim 100$ Myr, indicating that SFRs can fluctuate on a broad range of timescales. 
A similar conclusion was reached by \cite{2013MNRAS.434..209B} based on an analysis of the distribution of specific star formation rates as a function of stellar mass at $z\lesssim 0.3$. 
Both \cite{2012ApJ...744...44W} and \cite{2014MNRAS.441.2717K} find that in low-mass galaxies the amplitude of star formation bursts can be up to a factor of $\sim 30$. 
While some differences between SFRs inferred using different observational indicators can be explained in the context of steady star formation histories, such as due to incomplete sampling of the initial mass function (IMF), explanations based on constant SFRs tend to underestimate the magnitude of observed effects \citep[e.g.,][]{2009ApJ...706..599L, 2011ApJ...741L..26F}.

At higher redshift, \cite{2016ApJ...833...37G} find that the observed H$\beta$-to-UV ratio increases with $M_{\star}$ at $0.4<z<1$.\footnote{Since H$\beta$ is also powered by ionizing radiation, it probes the same timescale as H$\alpha$.}  
By comparing with lower-redshift measurements, \cite{2016ApJ...833...37G} also show that the H$\beta$-to-UV ratio decreases with increasing redshift. 
\cite{2016ApJ...833...37G} argue that their results are well explained by increasing star formation burstiness 
with both decreasing stellar mass and increasing redshift. 
\cite{2011ApJ...742..111V} identified an abundant population of $M_{\star}\sim10^{8}$ M$_{\odot}$  ``extreme emission line'' galaxies at $z\sim1.7$, which they interpret as having recently experienced intense starbursts of duration $\sim 15$ Myr \citep[for observations of extreme emission line galaxies at higher redshift, see][]{2017ApJ...838L..12F}. 
At $2.1<z<2.6$, \cite{2015ApJ...815...98S} measure larger scatter in the SFR-$M_{\star}$ relation for $M_{\star}\sim 10^{9.5}-10^{11.5}$ M$_{\odot}$ galaxies when H$\alpha$ is used to measure the SFR relative to the UV continuum (see also Smit et al. 2005\nocite{2015arXiv151108808S} for observational constraints at $z\sim4$). 
While \cite{2015ApJ...815...98S} caution that uncertainties in dust attenuation and IMF variations preclude directly interpreting this measurement in terms of bursty star formation, \cite{2017MNRAS.466...88S} compared the difference in the scatter between the simulated H$\alpha$- and UV continuum-derived SFR--$M_{\star}$ relations predicted by the FIRE simulations 
to the observations of \cite{2015ApJ...815...98S} and showed that the observed scatter difference is consistent with the order-of-magnitude SFR variations predicted by the simulations on timescales as short as a few Myr. 

Our model for bursty star formation builds on previous work on understanding the regulation of star formation by stellar feedback. 
Several studies have shown that simple analytic models in which ISM pressure sustained by stellar feedback (e.g., via turbulence) balances the weight of disc gas can explain the observed KS relation \citep[e.g.,][]{2005ApJ...630..167T, 2010ApJ...721..975O, 2011ApJ...731...41O, 2013MNRAS.433.1970F}. 
Recently, \cite{2017MNRAS.467.2301T} showed that such equilibrium models break down in galactic nuclei, where local dynamical timescales are too short for stellar feedback to effectively respond and establish a steady balance between feedback and gravity. 
\cite{2017MNRAS.467.2301T}'s simulations showed that the failure of stellar feedback to establish a steady equilibrium leads to bursty star formation in galactic nuclei. 
We analyze here other limits of feedback-regulated star formation models and argue that bursty star formation at high redshift and in dwarf galaxies can be understood as two different failures of stellar feedback to establish a steady equilibrium, the first due to the timescale for stellar feedback and the second due to the discreteness (stochastic sampling) of star-forming regions.  
At high redshift, where gas fractions are elevated, the two effects act in concert. 
Our derivations deliberately involve a number of simplifications and are intended to explain in simple terms \emph{why} star formation is bursty in some systems but time steady in others, rather than to be quantitatively exact. 
The fully dynamical numerical simulations referenced above are better suited for more detailed predictions.

We describe our star formation burstiness model and present our results in \S \ref{sec:model}. 
\S \ref{sec:discussion} discusses our findings and concludes. 
Appendices summarize supporting data and calculations. 
Throughout, we assume a standard flat $\Lambda$CDM cosmology with parameters consistent with the latest constraints \citep[$H_{0}\approx 70$ km/s/Mpc, $\Omega_{\rm m} = 1-\Omega_{\Lambda} \approx 0.27$ and $\Omega_{\rm b}\approx0.046$;][]{2016A&A...594A..13P}.
 
\section{Analytic model}
\label{sec:model}

\subsection{Preliminaries}
\label{sec:preliminaries}
Our model is based on several ideas for how star formation is regulated in galaxies. 
There are different models for galactic star formation in the literature but there is as yet no generally agreed upon theory. 
We therefore begin by briefly reviewing the key assumptions necessary to understand our model. 
This is not intended to be a thorough review of the field \citep[for reviews that also discuss alternate theories, see, e.g.,][]{2007ARA&A..45..565M, 2014PhR...539...49K} but rather to summarize the elements that we adopt in our modeling. 
We also note that our model for star formation burstiness is primarily based on recent simulations and analytic models of star formation regulation in galaxies. 
Although we review some observational constraints on star-forming regions for context below, we generally do not anchor our modeling to them; as we will explain, many results on individual star-forming regions are likely not representative of the physics that modulate galaxy-integrated SFRs.

\textbf{\textit{Most star formation occurs in Toomre-scale GBCs.}}
It is well known that stars form in molecular clouds \citep[e.g.,][]{1986ApJ...301..398M, 1988ApJ...334L..51M, 1989ApJ...339..149S, 1997ApJ...476..166W}. 
In this paper, we are interested in variability in the integrated SFR of galaxies, which is a sum over the SFRs of individual star-forming clouds. 
To correctly capture the expected SFR variance, it is critical to carefully define what counts as independent star-forming regions in the sum. 

The ISM is turbulent and highly inhomogeneous. 
As a result, clouds that undergo gravitational collapse generally have several different centers of collapse, each potentially corresponding to its own over-density of molecular gas and nascent star cluster. 
These multiple centers of collapse and resulting star clusters are evident in the filamentary images of GBC simulations \citep[e.g.,][]{2012ApJ...759L..27P, 2015ApJ...800...49L, 2016ApJ...829..130R, 2016arXiv161205635G}. 
The same simulations show that when the parent cloud collapses, a fraction $\epsilon_{\rm GBC}$ of the initial gas mass is converted into stars on a timescale of just $\approx (1-3)t_{\rm ff}^{\rm GBC}$, where $t_{\rm ff}^{\rm GBC}$ is the gravitational free fall time of the cloud evaluated at its mean initial density. 
When stellar feedback is included, it usually truncates star formation and limits $\epsilon_{\rm GBC}$. 
Since the different centers of collapse all form their stars during the collapse of the parent cloud, the parent cloud can be approximated as experiencing a single coherent burst on a timescale $\approx (1-3)t_{\rm ff}^{\rm GBC}$. 

If we are interested in galaxy-integrated burstiness, what distribution of independent star-forming clouds should we sum over?
Since parent GBCs can each contains a large number of molecular over-densities and form a large number of different star clusters, the usual molecular cloud or star cluster mass functions measured in observations \citep[e.g.,][]{2005PASP..117.1403R, 2012ApJ...752...96F, 2016ApJ...822...52R, 2017ApJ...834...57M} do not count the relevant independent star formation units but are instead expected to include correlated centers of collapse. 
We therefore appeal to theory to guide us. 

In general, galactic discs appear to be well-described by a Toomre $Q$ parameter near unity \citep[][]{1964ApJ...139.1217T, 1965MNRAS.130...97G}, corresponding to a state of marginal gravitational stability. 
In a smooth disc with $Q=1$, a single physical scale, the ``Toomre mass'' $M_{\rm T} \approx \pi h^{2} \Sigma_{\rm g}$ (where $h$ is the disc thickness and $\Sigma_{\rm g}$ is its gas surface density), is subject to gravitational instability. 
Perturbations on smaller length scales are stabilized by pressure in the disc, while perturbations on larger scales are stabilized by rotation. 
Smooth discs are therefore expected to fragment into GBCs of a definite mass scale $\sim M_{\rm T}$. 
During gravitational collapse, each Toomre-scale GBC would then hierarchically fragment into a large collection of smaller clouds, which together would correspond to the observed cloud mass function. 

Real galactic discs are not smooth but turbulent and in reality turbulent density fluctuations make a finite range of physical scales unstable to gravitational collapse. 
\cite{2013MNRAS.430.1653H} developed a general theory of gravitational fragmentation in turbulent media and addressed the ``clouds within clouds'' problem using an excursion set formalism (a generalization of the methodology used in cosmological structure formation to distinguish locally bound structures from larger structures that contain them; e.g. Bond et al. 1991\nocite{1991ApJ...379..440B}). 
In the excursion set terminology, the largest (parent) gravitationally bound structures correspond to ``first crossings'' of the collapse barrier. 
When applied to rotating discs, \cite{2013MNRAS.430.1653H} showed that the mass in first-crossing structures is strongly concentrated in objects near the Toomre scale (Hopkins 2013's Fig. 3). 

While the full mass function of gravitationally-bound clouds extends over many orders of magnitude, with a mass spectrum $dN/dm \sim m^{-\alpha}$ with $\alpha \sim 2$ in agreement with observational constraints on molecular cloud and star cluster mass functions, the first-crossing mass distribution decreases much more steeply with decreasing mass. The first-crossing mass distribution is instead strongly peaked near $M_{\rm T}$ (i.e., with most of the mass contained within just $\sim1-3$ orders of magnitude of $M_{\rm T}$). 
These first-crossing, parent, Toomre-scale GBCs undergoing coherent gravitational collapse are the independent clouds that we should sum over to properly capture the variance expected from stochastic sampling of star-forming regions.

\textbf{\textit{Within an individual GBC, star formation occurs in a burst.}} 
As mentioned above, dynamical simulations indicate that GBCs typically form stars in one main burst lasting $\approx (1-3)t_{\rm ff}^{\rm GBC}$. 
During this burst, the SFR in the GBC increases with time, until the collapse of the cloud is disrupted by stellar feedback. 
For example, the numerical simulations of \cite{2015ApJ...800...49L} indicate that $SFR \propto t$ in GBCs. 
At high redshift, where GBC masses tend to be large because of elevated gas fractions and where free fall times are shorter (see \S \ref{sec:stellar_evolution}-\ref{sec:stellar_evolution} below), this can lead to quite intense bursts in individual GBCs:
\begin{align}
\label{eq:SFR GBC burst}
SFR \sim 10~{\rm M_{\odot}~yr^{-1}}
\left( \frac{\epsilon_{\rm GBC}}{0.1} \right) 
\left( \frac{m}{\rm 10^{9}~M_{\odot}} \right) 
\left( \frac{t_{\rm ff}^{\rm GBC}}{\rm 10~Myr} \right).
\end{align}

We must address here observations that appear to contradict rapid, time-dependent in star formation in molecular clouds. 
Indeed, a large number of measurements indicate low star formation efficiencies per free fall time in molecular gas, $\epsilon_{\rm ff}^{\rm mol} \sim 0.01$, across large cloud samples and on different scales \citep[e.g.,][]{2007ApJ...654..304K, 2012A&A...539A...8G, 2012ApJ...745...69K, 2014ApJ...782..114E, 2016A&A...588A..29H}. 
Unfortunately, because the relevant units of coherent star formation in our model are Toomre-scale GBCs, it is generally not possible to directly compare star formation efficiencies measured for individual molecular clouds to our assumption of dynamic star formation in GBCs. 
This is because most observed star-forming clouds are not representative of Toomre-scale GBCs. Indeed, as explained above, most molecular clouds identified in observations are likely sub-units of more massive GBCs. 
Furthermore, most observed molecular clouds are not actually gravitationally bound. 
In the large molecular cloud catalog of \cite{2017ApJ...834...57M}, for example, only some of the most massive clouds are bound according to their measured virial parameter; by number most molecular clouds are unbound.

Nevertheless, observations do provide some support for dynamic star formation. 
Simulations with continuously driven turbulence predict that the star formation efficiency per free fall time 
is not universal but rather a function of the virial parameter of the cloud, $\alpha_{\rm vir}$, which measures the degree of gravitational boundedness via the ratio of kinetic energy to gravitational binding energy \citep[][]{1992ApJ...395..140B}. 
Dynamic star formation, on average, predicts an increasing star formation efficiency with decreasing virial parameter \citep[e.g.,][]{2012ApJ...759L..27P, 2015ApJ...800...49L}. 
A strong trend in qualitative agreement with this prediction has recently been observed in M51\citep{2017ApJ...846...71L}. 
In the Milky Way, attempts to identify a similar relationship with the virial parameter have however not revealed a clear trend \citep[e.g.,][]{2016ApJ...831...73V}. 
This suggests that the observational results may be sensitive to choices of observational tracers or definitions of the virial proxy or star formation efficiency (e.g., Leroy et al. studied regions of fixed size whereas Vutisalchavakul et al. analyzed clouds of varying size). 
Following a different approach, \cite{2016ApJ...833..229L} analyzed the dispersion of observationally-inferred $\epsilon_{\rm ff}^{\rm mol}$ in Milky Way molecular clouds and argued that the large dispersion is inconsistent with a time-independent $\epsilon_{\rm ff}^{\rm mol}$ and instead favors a time-variable efficiency.

Although clearly more work is needed test dynamic star formation observationally, we take the fact that dynamic star formation appears to be a generic prediction of simulations with self-gravity as our main motivation for assuming a time-dependent SFR in GBCs. 

\textbf{\textit{On galactic scales, star formation is regulated by a balance between gravity and stellar feedback.}} 
Averaged over entire galaxies, the star formation efficiency per free fall time, $\epsilon_{\rm ff}^{\rm gal}$, has low mean value $\sim0.02$ and relatively small dispersion 
\cite[e.g.,][]{1998ApJ...498..541K, 2010MNRAS.407.2091G, 2012ApJ...745...69K}. 
Our view in this paper is that the low star formation efficiency per free fall time on galactic scales is set by a global balance between the ISM pressure excited by stellar feedback and gravity, with only a weak (or no) dependence on the local star formation efficiency within individual GBCs. 
In this picture, $\epsilon_{\rm ff}^{\rm gal}$ is low because only a relatively small SFR is needed for stellar feedback to support the ISM against runaway gravitational collapse. 
Several analytic models have been formulated based on this ansatz and appear broadly consistent with observed star formation efficiencies \citep[][]{2005ApJ...630..167T, 2011ApJ...731...41O, 2013MNRAS.433.1970F}. 
Moreover, a number of hydrodynamic simulations including stellar feedback indicate that feedback can maintain the galactic ISM in rough vertical hydrostatic balance and/or that the star formation efficiency in a feedback-regulated ISM is not sensitive to the small-scale (sub-GBC) star formation prescription \citep[e.g.,][]{2011MNRAS.417..950H, 2012ApJ...754....2S, 2015ApJ...815...67K, 2017MNRAS.467.2301T, 2017arXiv170101788O}. 

Our picture of feedback-regulated star formation is distinct from another popular class of models in which it is assumed that the star formation efficiency in molecular clouds is universally low owing to the properties of supersonic turbulence in media with virial parameter of order unity \citep[e.g.,][]{2005ApJ...630..250K, 2009ApJ...699..850K, 2013MNRAS.436.3167F}. 
In those models, the low star formation efficiency on galactic scales is inherited from the low star formation efficiency $\epsilon_{\rm ff}^{\rm mol}\sim0.02$ in molecular clouds. 
In future work, it would be interesting to also investigate the galaxy-scale star formation variability predicted by universal star formation efficiency models, as this could provide a new test of the physics of star formation regulation.

\subsection{The FG13 feedback equilibrium model}
\label{sec:fg13}
The starting point for our analysis is the analytic feedback-regulated model of Faucher-Gigu\`ere, Quataert, \& Hopkins (2013; hereafter FG13)\nocite{2013MNRAS.433.1970F}. 
The FG13 model predicts a KS relation that emerges from a balance between gravity and supernova (SN) feedback in galactic discs, and also shows how the KS relation imposes certain consistency requirements on the number of massive star-forming GBCs active at any time.
We refer to that paper for more details and summarize here only the essential elements.  

We approximate galaxies with a two-zone model consisting of a volume-filling ISM and a collection of Toomre-scale GBCs in which star formation is confined.
The ISM is modeled as a thin disc with radius-dependent mean gas density $\bar{\rho}$ and is assumed to be in vertical hydrostatic balance, supported by turbulence excited by SN feedback. 
For simplicity we assume flat rotation curves, which we model using an isothermal potential with velocity dispersion $\sigma$ \citep[see also][]{2005ApJ...630..167T}. 
The circular velocity is then $v_{\rm c} = \sqrt{2} \sigma$. 

Throughout this paper, we assume that galactic discs tend to self-regulate to a Toomre 
parameter
\begin{equation}
\label{Q def}
Q = \frac{\kappa c_{\rm T}}{\pi G \Sigma_{\rm g}}  = \frac{2 \sigma c_{\rm T}}{\pi G \Sigma_{\rm g} r} \approx 1,
\end{equation}
where $\kappa\equiv \sqrt{4 \Omega^{2} + d\Omega^{2}/d\ln{r}} = 2 \sigma / r$ is the epicyclic frequency
and $c_{\rm T}$ is the velocity dispersion of the turbulence \citep[e.g.,][]{1972ApJ...176L...9Q, 1989ApJ...344..685K, 2001ApJ...555..301M}. 
We expect the ISM to self-regulate to this value, corresponding to marginal gravitational stability, because  turbulence dissipation tends to decrease $Q$ below unity. 
When $Q < 1$, gravitational fragmentation increases the star formation rate and the associated heating of the disc by stellar feedback pushes the disc back to $Q\approx1$. 
Using $\Sigma_{\rm g} \equiv 2 h \bar{\rho}$, where $h$ is the gaseous disc scale height, we can solve for the  free fall time in the ISM at the half-mass radius $r_{\rm 1/2}$, 
\begin{align}
\label{eq:tff disc}
t_{\rm ff}^{\rm disc}(r_{\rm 1/2}) &= \left( \frac{3 \pi}{32 G \bar{\rho} (r_{\rm 1/2})} \right)^{1/2} \\ \notag
& = \left( \frac{3 Q}{64\times 2^{1/2}} \right)^{1/2} t_{\rm orb}(r_{\rm 1/2}),
\end{align}
where $t_{\rm orb}(r) \equiv 2 \pi r / v_{\rm c}$ is the orbital time. 
For $Q=1$, the free fall time reduces to a constant fraction of the orbital time: 
\begin{align}
\label{eq:tff disc Q1}
t_{\rm ff}^{\rm disc}(r_{1/2}) & \approx 0.2 t_{\rm orb}(r_{1/2}).
\end{align}

By balancing the turbulent pressure with the weight of the gaseous disc normal to the disc plane, FG13 derived the following expression for the KS relation:
\begin{align}
\label{star formation law momentum balance}
\dot{\Sigma}_{\star} & = \frac{2 \sqrt{2} \pi G Q }{\mathcal{F}} \left( \frac{P_{\star}}{m_{\star}} \right)^{-1} \Sigma_{\rm g}^{2}.
\end{align}
Here, $\dot{\Sigma}_{\star}$ is the star formation rate surface density, $P_{\star}/m_{\star}$ is the momentum injected in turbulence by SN feedback per stellar mass formed, 
and $\mathcal{F}$ encapsulates uncertain factors of order unity. 
FG13 showed that $\mathcal{F}=2$ provides a good fit to observations for $P_{\star}/m_{\star} \approx 3,000$ km/s, appropriate for SN feedback and a standard Kroupa IMF \citep[e.g.,][]{1988ApJ...334..252C, 2011ApJ...731...41O, 2015MNRAS.450..504M}. 
FG13 also showed that the prediction of this model can be expressed in terms of the dimensionless star formation efficiency per free fall time on galactic scales (such that $\dot{\Sigma}_{\star} \equiv \epsilon_{\rm ff}^{\rm gal} \Sigma_{\rm g}/t_{\rm ff}^{\rm disc}$) as
\begin{align}
\label{eps ff gal alpha infty}
\epsilon_{\rm ff}^{\rm gal} \approx \frac{\sqrt{3} \pi}{2^{7/4} \mathcal{F}} \frac{f_{\rm g} v_{\rm c}}{P_{\star} / m_{\star}},
\end{align}
where $f_{\rm g}$ is the gas mass fraction. 
For most observed galaxies, this galaxy-scale star formation efficiency is small, of order one to a few percent \citep[e.g.,][]{2010MNRAS.407.2091G, 2012ApJ...745...69K}. 
Starbursts can be induced in galaxy mergers as a result of strong gravitational torques that efficiently funnel large amounts of gas into the nucleus of the merger remnant \citep[e.g.,][]{1972ApJ...178..623T, 1991ApJ...370L..65B}. 
In this case, the burst can occur on a short timescale because the high-density gas concentration in the nucleus can reach short dynamical times $\lesssim 1$ Myr in the inner $100$ pc. 
However, galaxy mergers are too rare to explain the frequent SFR fluctuations seen in Figure \ref{fig:FIRE_SFHs_norm}. 

In the following sections, we describe two regimes in which ordinary (non-merging) galaxies can experience large deviations from the KS relation. 
These deviations and the accompanying bursts of star formation can occur either spontaneously due to instabilities in the disc or as a result more minor external perturbations, such as smooth inflows from the intergalactic medium. 

\begin{figure}
\begin{center}
\includegraphics[width=0.475\textwidth]{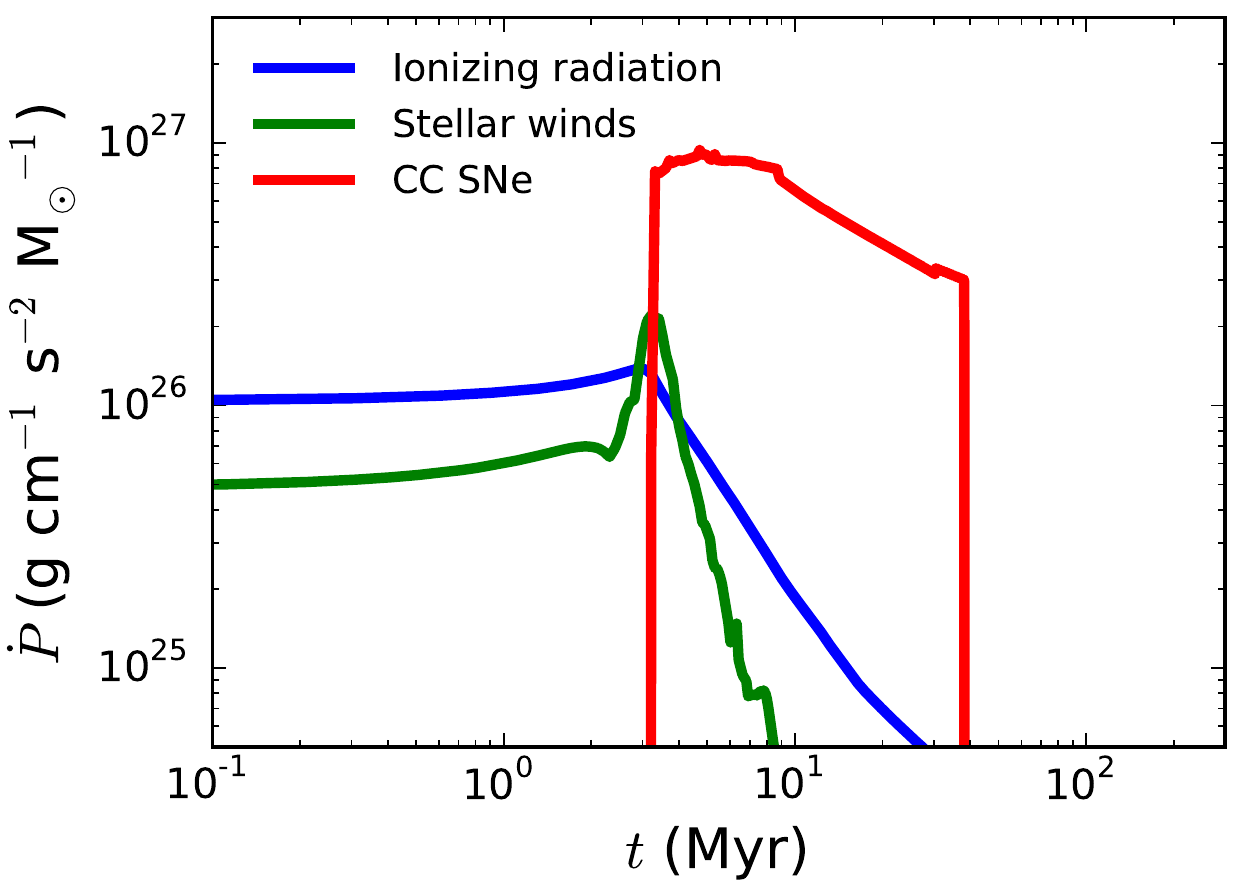}
\end{center}
\caption[]{Momentum flux output by the main stellar feedback processes (ionizing radiation, stellar winds, and core collapse supernovae) per stellar mass formed as a function of time since a burst of star formation (see \S \ref{sec:stellar_evolution} for details). 
Prompt feedback from ionizing radiation and stellar winds can disrupt stellar birth clouds but the integrated momentum output in the ISM is dominated by SNe.
}
\label{fig:SB99_Pdot} 
\end{figure}
 
\label{sec:SB99_Pdot}

\subsection{Breakdown of equilibrium due to the supernova feedback timescale}
\label{sec:stellar_evolution}
Implicit in FG13's equilibrium model is the assumption that SN feedback can respond sufficiently rapidly to gravitational collapse of the disc to establish hydrostatic balance. 
\cite{2017MNRAS.467.2301T} noted that this assumption breaks down in galactic nuclei, where local dynamical timescales are shorter than a ``stellar feedback'' timescale. 
We argue here that a similar effect is in part responsible for bursty star formation in high-redshift galaxies (as a whole). 
In the early Universe the characteristic dynamical timescales of galaxies were shorter, but stellar evolution proceeded at a constant rate. 
As the disc free fall time becomes smaller than the stellar feedback timescale, it becomes impossible for the bulk of galaxies to reach a tight balance between feedback and gravity. 
In this limit, the instantaneous star formation efficiency in the disc can deviate from the median KS relation by a large factor. 

Figure \ref{fig:SB99_Pdot} shows the momentum flux output by the main stellar feedback processes (ionizing radiation, stellar winds, and core collapse supernovae) per stellar mass formed as a function of time since a burst of star formation. 
The plot, produced using STARBURST99 v7.0.1 \citep[][]{1999ApJS..123....3L}, assumes a \cite{2001MNRAS.322..231K} IMF and solar metallicity. 
The SN energy rate returned by STARBURST99 is for the ``prompt'' kinetic energy of SN ejecta. 
As SNe expand into the ISM, the radial momentum of SN remnants is enhanced by about an order of magnitude during the Sedov-Taylor (energy-conserving) phase \citep[e.g.,][]{1988ApJ...334..252C, 1998ApJ...500..342B, 2015MNRAS.450..504M}. 
This ``boosted'' momentum is the momentum available to drive interstellar turbulence and galactic winds. 
We therefore plot in the figure the boosted SN momentum calculated using fits to the evolution of SNRs in an inhomogeneous ISM from the simulations of \cite{2015MNRAS.450..504M}.\footnote{Specifically, we use the fit  appropriate for an inhomogeneous ISM with mean density $\langle n_{\rm H} \rangle = 1$ cm$^{-3}$, Mach number $\mathcal{M}=30$, and solar metallicity. Each supernova is assumed to have an energy of $10^{51}$ erg.} 
As the figure shows, SNe dominate the momentum input in the ISM by a large factor relative to radiation and stellar winds. 

SN feedback has two characteristic timescales. 
The first is the timescale ($t_{\rm 1st}\approx 3$ Myr) for the first SNe to explode following a burst of star formation. 
Before this time, SN feedback cannot oppose the gravitational collapse converting gas into stars. 
The second is the timescale for most SNe to explode. 
This is roughly the timescale over which SN feedback acts. 
For a Kroupa IMF, $\approx60\%$ of the cumulative energy and momentum from core collapse SNe have been returned in the ISM by $\approx 20$ Myr following a burst of star formation (insensitive to metallicity).
We thus assume a SN feedback timescale $t_{\rm SN}\approx 20$ Myr for the numerical estimates in this paper. 

The characteristic timescales of SN feedback introduce two different effects. 
In a medium with $t_{\rm ff} \ll t_{\rm 1st}$, most of the gas can be converted into stars in a rapid burst before SN feedback can disperse the birth cloud. 
As we will show below, typical free fall times are particularly short in the early Universe so this limit could be relevant for explaining the early formation of bound star clusters that are observed as old globular clusters today (which requires a high star formation efficiency). 
Properly understanding this limit would require modeling the effects of radiative feedback which acts before $t_{\rm 1st}$, which is beyond the scope of this paper. 
We focus instead on effects introduced by the longer timescale $t_{\rm SN}$. 
Since SNe dominate the total momentum output into the ISM, we assume that SN feedback is the primary source of turbulent pressure in the ISM.\footnote{There may be exceptions, such as at very high gas surface densities where multiple scatterings of re-processed infrared radiation can enhance radiative feedback \citep[][]{2003JKAS...36..167S, 2005ApJ...630..167T} or in low surface density regions, where photoionization and photoheating can be important \citep[e.g.,][]{2010ApJ...721..975O}.}

When $t_{\rm ff}^{\rm disc} \lesssim t_{\rm SN}$, a feedback-supported galactic disc 
will be susceptible to large deviations from the KS relation. 

Initially, this is because gravitational collapse on a timescale $\approx t_{\rm ff}^{\rm disc}$ can form stars with an elevated efficiency before SN feedback can respond (on a timescale $\approx t_{\rm SN}$) and push the disc back toward vertical hydrostatic balance with $Q\approx1$. 
Subsequently, an overshoot effect can occur. 
This is because stellar feedback continues to operate for a fixed period of time (set by stellar evolution) after star formation ends locally, regardless of the state of the surrounding gas, which can lead to strong gas blowouts. 
Such bursts of outflowing gas are seen in the galactic nucleus simulations of \cite{2017MNRAS.467.2301T}. 
They are also seen as ``gusty'' galactic winds following star formation bursts in cosmological simulations such as the FIRE simulations shown in Figure \ref{fig:FIRE_SFHs_norm}. 
The near evacuation of the ISM by strong outflows explains the periods of highly suppressed star formation apparent in Figure \ref{fig:FIRE_SFHs_norm}, where blowouts can suppress the instantaneous SFR by up to $\gtrsim2$ orders of magnitude.

We show below that the limit of unstable feedback regulation described above, leading to star formation burst-outflow-suppressed star formation cycles,  is typical of high-redshift ($z \gtrsim 1$) galaxies of all masses and of gas-rich dwarf galaxies all the way to the present time. 

For the free fall time, we use a simple model for the cosmological evolution of galactic discs, inspired by classic models in which disc sizes can be predicted by assuming that baryons have the same specific angular momentum as their parent dark matter halos \citep[e.g.,][]{1980MNRAS.193..189F, 1997ApJ...482..659D, 1998MNRAS.295..319M}. 
Specifically, we assume that at any redshift the half-mass disc radius is a constant fraction $\approx 2\%$ of the halo virial radius:
\begin{equation}
\label{eq:shibuya_scaling}
r_{1/2} \approx 0.02 R_{\rm vir}. 
\end{equation} 
\cite{2015ApJS..219...15S} shows that this scaling holds in observations (on average to better than a factor of 2) from $z=0$ to $z\sim 8$. 
In a study focused at low redshift, \cite{2013ApJ...764L..31K} found that galaxies obey a consistent scaling over more than eight orders of magnitude in stellar mass. 
Here, $R_{\rm vir}$ is the virial radius for halos defined to have a redshift-dependent enclosed overdensity $\Delta_{\rm c}(z)$ relative to the critical density, $\rho_{\rm c}$, following \cite{1998ApJ...495...80B}. 
\S \ref{sec:disc_radii} in the Appendix provides more details on how we compute galaxy radii. \begin{figure}
\begin{center}
\includegraphics[width=0.475\textwidth]{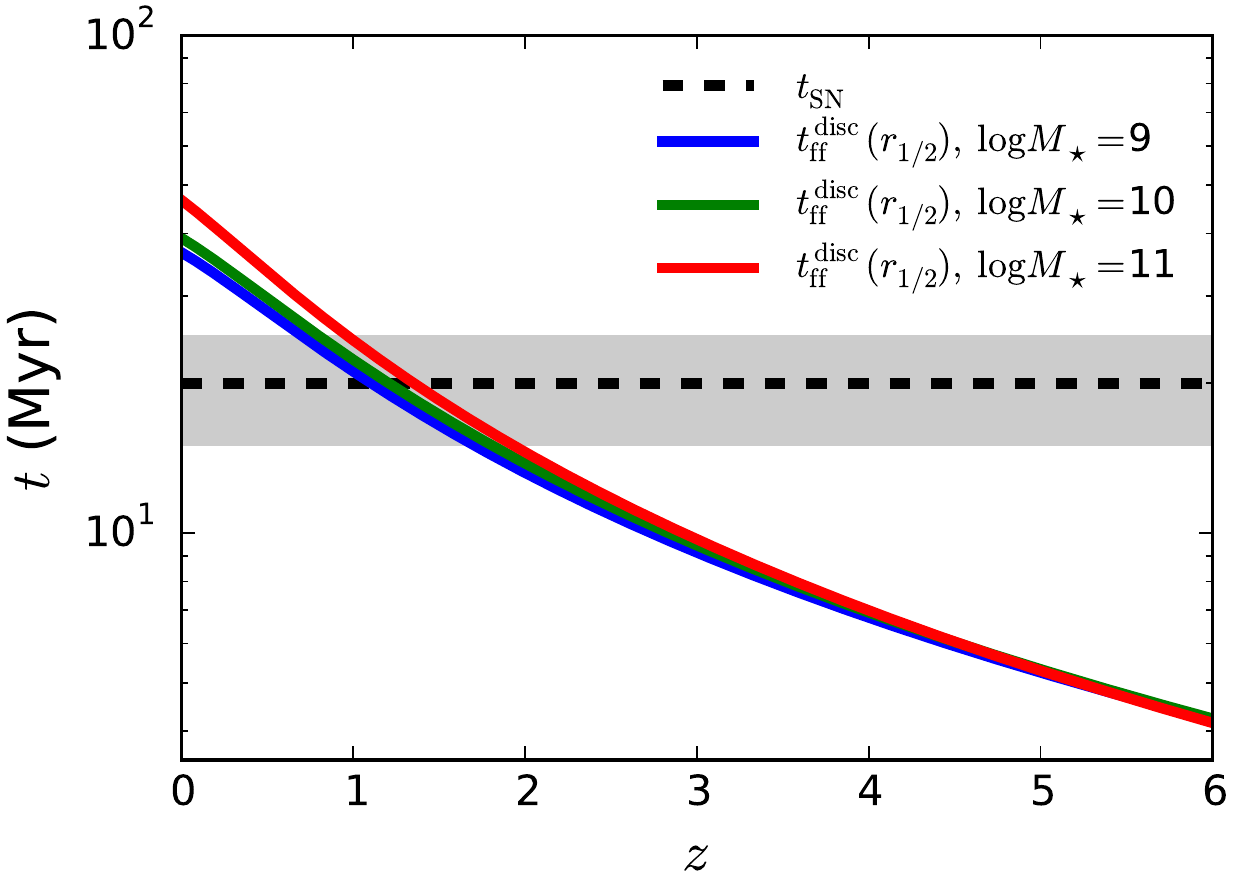}
\end{center}
\caption[]{Comparison of the free fall time at the half-mass radius of galactic discs, $t_{\rm ff}^{\rm disc}$, as a function of redshift for different stellar masses (solid colored curves) to the supernova feedback timescale $t_{\rm SN}=20$ Myr (over which most of the SN feedback acts; dashed black) as a function of redshift. 
Galactic dynamical times are a constant fraction of the age of the Universe, so they are shorter at high redshift (only weakly dependent on galaxy mass), while the SN feedback timescale is constant. 
At $z\gtrsim1$, characteristic free fall times become shorter than the SN feedback timescale and the feedback cannot effectively respond to gravitational collapse. 
Galactic star formation should be bursty in this regime
The grey band covers $\pm 5$ Myr around the fiducial feedback timescale as an indication of the sensitivity of the results to the particular choice.
}
\label{fig:tff_vs_star} 
\end{figure}

We note that cosmological hydrodynamic simulations clearly show that feedback (and not just angular momentum inherited from halo formation) is critical to produce realistic galaxy sizes \citep[e.g.,][]{2010MNRAS.409.1541S, 2015MNRAS.450.1937C, 2016ApJ...824...79A}. 
We do not attempt here to model in detail what sets galaxy sizes, but rather simply use the above observationally-supported scaling. 
We focus in this paper on late-type star-forming galaxies. 
Early-type galaxies, which we do not consider, may retain a smaller fraction of the specific angular momentum of their parent halos and may therefore be somewhat more compact \citep[e.g.,][]{2015ApJ...804L..40G}. 

Next, we need a model for the circular velocities of galaxies. 
In \S \ref{sec:TF} of the Appendix, we show that a model in which galaxy circular velocities are a constant factor of the maximum circular velocity of NFW \citep{1997ApJ...490..493N} halos simultaneously matches the observed Tully-Fisher relation at $z=0$ and its evolution to $z=2$. 
In this model, the galaxy circular velocity is related to halo properties following
\begin{align}
v_{\rm c} \approx 0.465 \left( \frac{c_{\rm vir}}{A(c_{\rm vir})} \right)^{1/2} \left( \frac{G M_{\rm vir}}{R_{\rm vir}} \right)^{1/2},
\end{align}
where $c_{\rm vir}$ is the halo concentration (see eq. \ref{eq:concentration}) and $A$ is a dimensionless function of concentration (see eq. \ref{eq:A}). 
We use the \cite{2013MNRAS.428.3121M} abundance matching relation to convert between halo mass and galaxy stellar mass. 

For a disc with $Q=1$, we can use equation (\ref{eq:tff disc Q1}) to evaluate the free fall time in the ISM:
\begin{align}
t_{\rm ff}(r_{1/2}) &\approx 0.0264 \left( \frac{1}{G \rho_{\rm c}(z)} \right)^{1/2} g(z) \\ \notag
& \approx \frac{0.0764}{H(z)} g(z),
\end{align}
where 
\begin{equation}
g(M_{\rm vir},~z) \equiv \left( \frac{A(c_{\rm vir})}{\Delta_{\rm c}(z) c_{\rm vir}}\right)^{1/2}
\end{equation}
and $H$ is the Hubble parameter. 
In our numerical calculations, we evaluate the function $g$ using the full equations given in the Appendix but note that it is only weakly dependent on mass and redshift. 

Figure \ref{fig:tff_vs_star} compares $t_{\rm ff}^{\rm disc}(r_{\rm 1/2})$ for different fixed stellar masses and $t_{\rm SN}$ versus redshift. 
For $z \lesssim 1.3$, $t_{\rm SN} < t_{\rm ff}^{\rm disc}(r_{\rm 1/2})$ and SN feedback has enough time to establish an equilibrium and effectively regulate star formation in galaxies. 
At $z \gtrsim 1.3$, however, the feedback timescale becomes longer than $t_{\rm ff}^{\rm disc}(r_{\rm 1/2})$ so galaxies can experience intense bursts of star formation to which SN feedback cannot respond sufficiently rapidly.\footnote{Another relevant timescale is the time necessary for pre-existing ISM turbulence to dissipate (absent a driving source), $t_{\rm turb}$. If $t_{\rm turb}$ were longer than $t_{\rm ff}^{\rm disc}$, turbulent pressure support would prevent the disc from collapsing on a free fall time and our argument would require modification. However, we can show that $t_{\rm turb}\lesssim t_{\rm ff}^{\rm disc}$ in general. 
Assuming that the largest turbulent eddies have a size equal to the disc thickness $h$, $t_{\rm turb}\approx h / c_{\rm T}$ \citep[this is true for both supersonic and subsonic turbulence, and also in the presence of magnetic fields; e.g.][]{1998ApJ...508L..99S, 1999ApJ...524..169M}. Using $h/r \approx c_{\rm T}/v_{\rm c}$ (FG13) to eliminate $h$ and $c_{\rm T}$ in favor of $r$ and $v_{\rm c}$, we find that for a disc with $Q\approx 1$, $t_{\rm turb} \approx 0.16 t_{\rm orb} \lesssim t_{\rm ff}^{\rm disc}$ (see eq. \ref{eq:tff disc Q1}), independent of other disc properties.} 
This result is nearly independent of galaxy mass and we propose that it explains in part why high-resolution cosmological simulations with time-resolved stellar feedback show bursty star formation histories at high redshift for all galaxies \citep[e.g.,][]{2014MNRAS.445..581H, 2017MNRAS.470.1050F}. 
Quantitatively, this result is in good agreement with the transition from bursty to time steady star formation occurring around $z\sim1$ for massive galaxies in the FIRE simulations (Fig. \ref{fig:FIRE_SFHs_norm}).

Since $g$ is only weakly dependent on $M_{\rm vir}$ and $z$, we can push our analytics further to derive an expression for the ``burstiness redshift'' $z_{\rm burst}$ above which we expect bursty star formation. 
For this estimate, we use $g\approx0.034$. 
Then,
\begin{equation}
t_{\rm ff}^{\rm disc}(r_{1/2}) \approx \frac{0.0026}{H(z)}.
\end{equation}
This is a version applied to galactic discs of the well-known result that the characteristic dynamical time of halos is a constant fraction of $1/H(z)$. 
Since the age of the Universe $\sim 1/H(z)$ at any redshift, this implies that the characteristic free fall times of galactic discs must become shorter than the constant $t_{\rm SN}$ at high redshift. 

At $z\gtrsim 1$, where dark energy is negligible, $H(z)\approx H_{0}\sqrt{\Omega_{\rm m} (1+z)^{3}}$. 
We can then analytically solve for $z_{\rm burst}$ by setting $t_{\rm SN}=t_{\rm ff}^{\rm disc}(r_{\rm 1/2})$:
\begin{align}
z_{\rm burst} & \approx \left( \frac{0.0026}{H_{0} t_{\rm SN} \sqrt{\Omega_{\rm m}}} \right)^{2/3} - 1 \\ \notag
& \approx 2.3 \left( \frac{t_{\rm SN}}{\rm 20 Myr} \right)^{-2/3} - 1.
\end{align}
This matches the nearly mass-independent crossing of $t_{\rm SN}$ and $t_{\rm ff}^{\rm disc}(r_{1/2})$ in Figure \ref{fig:tff_vs_star}. 
The transition from bursty to time steady will in practice be gradual around that redshift because feedback and gravitational collapse both operate continuously in time, as well as because galaxies have significant dispersion in their properties at any mass and redshift (which affect their internal free fall times). 
We indicate this in the figure with the horizontal grey band, which covers $\pm 5$ Myr around the fiducial feedback timescale as an approximation of the sensitivity of our results to the choice of feedback timescale.

\subsection{Breakdown of equilibrium due to the discreteness of star formation}
\label{sec:discreteness}
In the FIRE simulations shown in Figure \ref{fig:FIRE_SFHs_norm}, massive galaxies settle into a more time steady mode of star formation at low redshift 
but dwarf galaxies continue to experience bursty star formation down to $z\sim 0$. 
The burstiness of dwarfs at $z<1$ cannot be explained by the previous galaxy-scale timescale argument, but we propose that it can instead be explained by the discreteness of star-forming regions. 

As discussed in \S \ref{sec:preliminaries}, most of the star formation in galaxies at any given time occurs in Toomre-scale GBCs. In individual GBCs, the simulations summarized in \S \ref{sec:preliminaries} show that the SFR tends to be bursty as the cloud collapses. 
In the limit in which only a small number of GBCs contribute to the galactic SFR at any given time, the galactic SFR will inherit the time dependence of GBC-scale star formation. 
In nearby galaxies in which the KS relation has been studied at high spatial resolution, the scatter is measured to increase by a large factor when the relation is measured for galaxy patches of size $\lesssim300$ pc-$1$ kpc \citep[][]{2010ApJ...722.1699S, 2011ApJ...735...63L} rather than averaging over the entire galaxy. 
These observations are consistent with stochastic sampling of bright star-forming regions causing departures from the median KS relation, since fewer star-forming regions are averaged over when measuring on small scales within galaxies \citep[e.g.,][]{2017MNRAS.467.2301T, 2017arXiv170101788O}. 
In analogy with the averaging necessary for the KS relation to be tight, we expect that only galaxies in which a sufficiently large number of GBCs efficiently form stars at any time can produce time-steady galactic SFRs.

In what follows, we derive a scaling for how the number of Toomre-scale GBCs depends on galaxy properties. 
Since galaxies contain GBCs spanning a spectrum of masses but most star formation is expected to occur within coherently collapsing Toomre-scale clouds, we characterize galaxies by the parameter $N_{\rm U}$, defined as the number of GBCs that would be present if all of the self-gravitating mass were distributed in clouds of mass $M_{\rm U} \approx M_{\rm T}$, the maximum (upper) GBC mass. 
Appendix \ref{sec:SFR_var_vs_NGMC} shows how 
the minimum normalized SFR variance, $\sigma_{\rm SFR}/\langle SFR \rangle$, scales with $N_{\rm U}$. 
Specifically, 
for individual GBCs with $SFR \propto t$,
\begin{align}
\label{eq:SFR_poisson_main}
\frac{\sigma_{\rm SFR}}{\langle SFR \rangle} \approx \frac{1}{\sqrt{\gamma N_{\rm U}/0.5}},
\end{align}
where $\gamma$ is the fraction of the time each GBC actually forms stars. 
As shown in the appendix, the exact numerical value of the numerator in the above expression depends on the parameters describing the GBC mass function, but only weakly. 
This result shows that the SFR variance is roughly the Poisson variance expected for the number of Toomre-scale GBCs actively forming stars at any given time. 
Equation (\ref{eq:SFR_poisson_main}) is the minimum SFR variance expected because it only takes into account stochastic sampling of Toomre-scale regions. 
In particular, this estimate neglects the near-complete SFR suppression that can follow star formation bursts owing to evacuation of the gas reservoir (see \S \ref{sec:stellar_evolution}), or temporary enhancements due to accretion of new gas (either smoothly from the intergalactic medium or in galaxy mergers). 
These effects can significantly increase the actual SFR variance that will be observed in bursty galaxies.

\begin{figure}
\begin{center}
\mbox{
\includegraphics[width=0.475\textwidth]{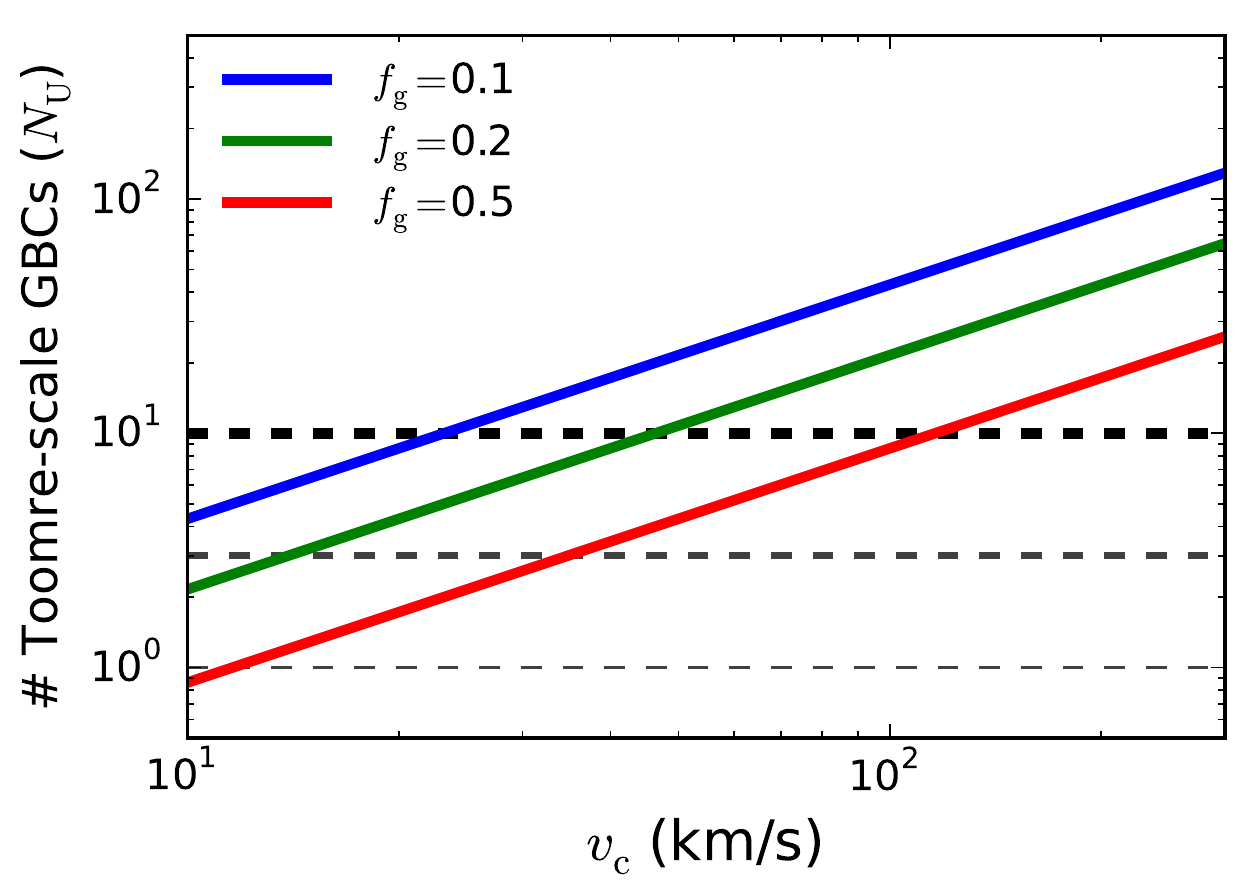}
}
\end{center}
\caption[]{Number of Toomre-scale GBCs as a function of galaxy circular velocity, for different gas mass fractions indicated by the different colors (the curves do not depend on redshift). 
Within a single GBC, star formation can be highly time-dependent as gravitational collapse proceeds, so a large number of GBCs must contribute in order for the galactic average to be time steady. 
The horizontal dashed lines indicate different values $N_{\rm U}=1,~3,~10$ below which we expect increased SFR variability (\S \ref{sec:discreteness} quantifies the minimum SFR variance expected vs. $N_{\rm U}$). 
Burstiness is predicted to increase in dwarf galaxies with low $v_{\rm c}$. 
Burstiness should also increase with increasing gas fraction 
at fixed stellar mass because the Toomre mass increases strongly with $f_{\rm g}$, so that the total number of Toomre-scale GBCs decreases.}
\label{fig:N_GMC_vs_vc} 
\end{figure}

\begin{figure*}
\begin{center}
\mbox{
\includegraphics[width=0.49\textwidth]{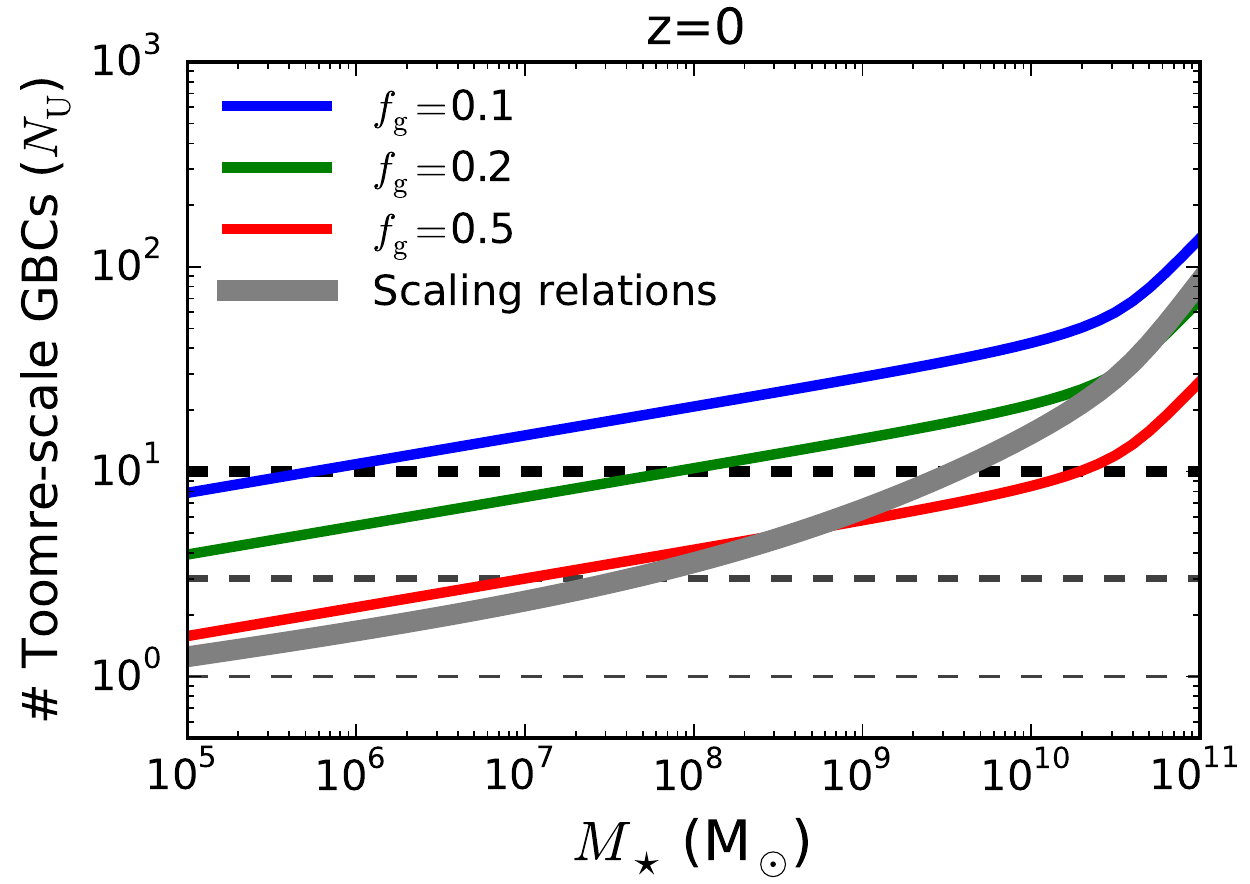}
\includegraphics[width=0.49\textwidth]{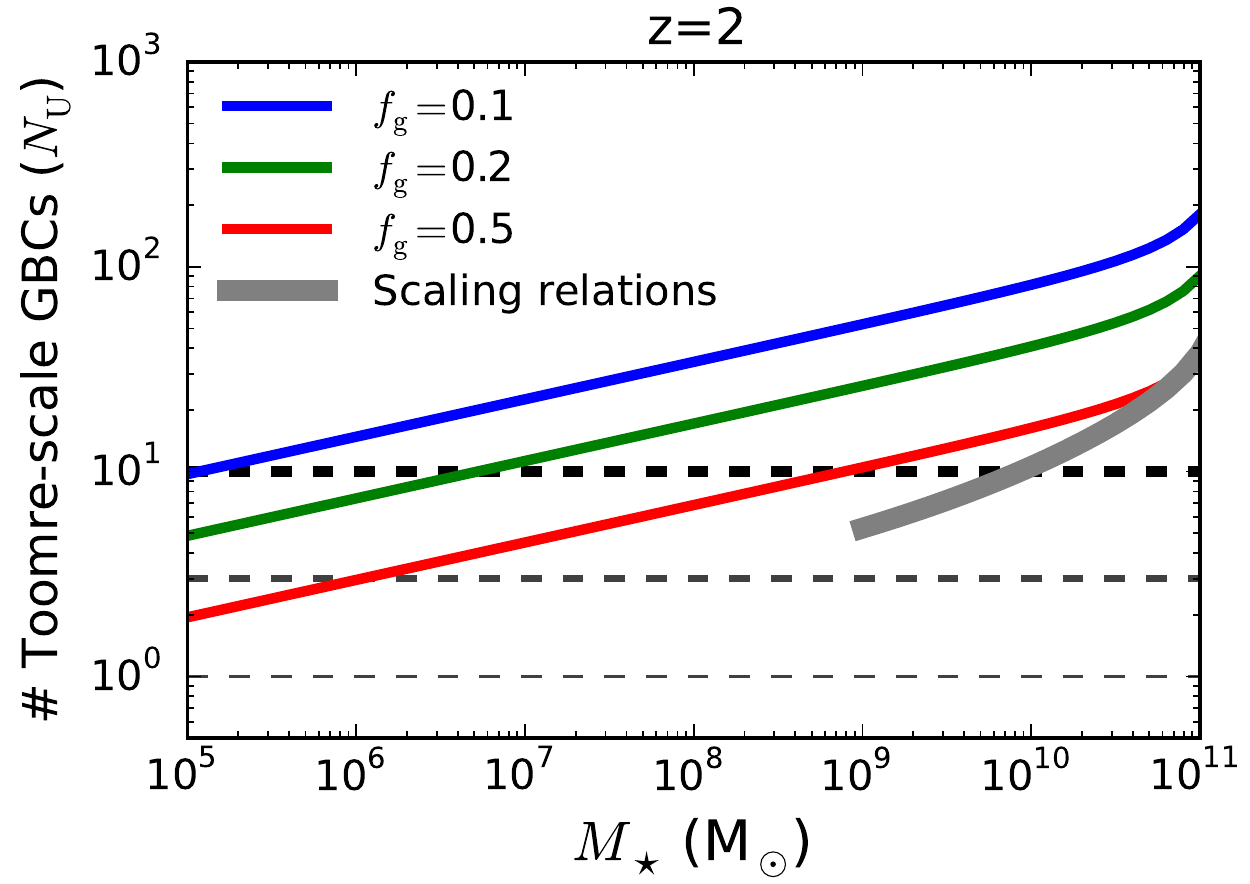}
}
\end{center}
\caption[]{Number of Toomre-scale GBCs (as in Fig. \ref{fig:N_GMC_vs_vc}) as a function of galaxy stellar mass. 
These curves depend on redshift because the $M_{\star}-v_{\rm c}$ relation is redshift dependent. 
In each panel, the solid colored curves show predictions for fixed gas fractions and the thick grey curve shows the prediction for the typical gas fraction for the given galaxy mass and redshift using the scaling relations summarized in Appendix \ref{sec:galaxy_scalings}. 
The horizontal dashed lines indicate different values $N_{\rm U}=1,~3,~10$ below which we expect increased SFR variability (\S \ref{sec:discreteness} quantifies the minimum SFR variance expected vs. $N_{\rm U}$). 
At $z=2$, galaxies of a given stellar mass are predicted to be more bursty (fewer massive GBCs) because of their enhanced gas fractions. 
This effect is in addition to the burstiness expected at high redshift from the SN feedback timescale being longer than the characteristic ISM free fall times (Fig. \ref{fig:tff_vs_star}).
}
\label{fig:N_GMC_vs_Mstar} 
\end{figure*}

Assuming that all star formation proceeds in GBCs and that each GBC ultimately converts a fraction $\epsilon_{\rm int}^{\rm GBC}$ of its initial gas mass into stars (determined by how stellar feedback operates in GBCs), the requirement that galaxies lie on the KS relation constrains the fraction $f_{\rm GBC}$ of the total ISM mass found in GBCs at any time. 
Using $M_{\rm U} \approx M_{\rm T} \approx \pi h^{2} \Sigma_{\rm g}$, 
we find that
\begin{align}
N_{\rm U} &= \frac{2 ( \pi r_{1/2}^{2}) \Sigma_{\rm g}(r_{1/2}) f_{\rm GBC}}{M_{\rm U}} \\ \notag
& = 2 \left( \frac{r_{1/2}}{h} \right)^{2} f_{\rm GBC} \\ \notag
& = \frac{16}{Q^{2} f_{\rm g}^{2}} f_{\rm GBC}.
\end{align}
The last expression uses the fact that for our disc model $h/r = Q f_{\rm g} / 2^{3/2}$ (i.e., more gas rich discs are thicker) to eliminate the disc thickness in favor of the gas fraction. 
Using this relation, we can also show that 
$M_{\rm U} \propto f_{\rm g}^{3} M_{\star}$. 
We note that, because the SFR in individual GBCs is time dependent, only a fraction of all GBCs will be observed as luminous star-forming regions at any given time, consistent with the wide range of instantaneous star formation efficiencies inferred in Milky Way GBCs \citep[e.g.,][]{2016ApJ...833..229L}. 

To make progress, we need an expression for $f_{\rm GBC}$. 
FG13 showed that since we assume that all star formation occurs in GBCs, we can write
\begin{align}
\label{eta consistency}
\epsilon_{\rm ff}^{\rm gal} = \frac{f_{\rm GBC}}{\tilde{t}_{\rm GBC}} \epsilon_{\rm int}^{\rm GBC},
\end{align}
where
\begin{align}
\label{tilde t GMC}
\tilde{t}_{\rm GBC} \equiv \left( \frac{t_{\rm GBC}}{t_{\rm ff}^{\rm disc}} \right)
\end{align}
and $t_{\rm GBC}$ is the lifetime of GBCs. 
This is simply the consistency condition that the sum of star formation occurring in all GBCs must equal the total galactic star formation. 
In what follows, we assume $\tilde{t}_{\rm GBC} \approx 1$. 
This is a good approximation to the GBC simulations discussed in \S \ref{sec:preliminaries}, noting that GBCs assemble on a timescale $t_{\rm ff}^{\rm disc}$, and models show that GBCs collapse and get dispersed by stellar feedback on a timescale comparable to their free fall time \citep[e.g.,][]{2016ApJ...819..137K, 2016arXiv161205635G}.
Then
\begin{align}
f_{\rm GBC} & \approx \frac{\epsilon_{\rm ff}^{\rm gal}}{\epsilon_{\rm int}^{\rm GBC}}
\end{align}
and therefore, using equation (\ref{eps ff gal alpha infty}) for $\epsilon_{\rm ff}^{\rm gal}$,
\begin{align}
N_{\rm U} \approx \frac{8 \sqrt{3} \pi v_{\rm c}}{2^{3/4} \mathcal{F} Q^{2} f_{\rm g} (P_{\star}/m_{\star}) \epsilon_{\rm int}^{\rm GBC}}.
\end{align} 
The model thus predicts decreasing $N_{\rm U}$, and therefore increasing SFR variability, with increasing gas fraction (at fixed $v_{\rm c}$). 

Defining a critical minimum number of Toomre-scale GBCs necessary to exhibit steady star formation, $N_{\rm U}^{\rm crit}$, we can a derive a minimum galaxy circular velocity necessary for steady star formation:
\begin{align}
v_{\rm c,burst}  & \approx \frac{2^{3/4} \mathcal{F} Q^{2} (P_{\star}/m_{\star}) \epsilon_{\rm int}^{\rm GBC} f_{\rm g}}{8 \sqrt{3} \pi} N_{\rm U}^{\rm crit} \\ \notag
& \approx 46~{\rm km/s} \left( \frac{f_{\rm g}}{0.2} \right)  \left( \frac{P_{\star}/m_{\star}}{\rm 3,000~km/s} \right) \\ \notag
& ~~~~~~~~~~~~~~~~~~~~~~~~~~~~~~~\times \left( \frac{\epsilon_{\rm int}^{\rm GBC}}{0.1} \right)\left( \frac{N_{\rm U}^{\rm crit}}{10} \right),
\end{align}
where the last numerical expression assumes $Q=1$, 
$\mathcal{F}=2$, and $P_{\star}/m_{\star}=3,000$ km/s as before. 
Figure \ref{fig:N_GMC_vs_vc} plots $N_{\rm U}$ versus $v_{\rm c}$ for different gas fractions. 
We note that in reality there is no single critical threshold below which galaxies will be bursty; rather that there will be a gradual transition of increasing burstiness with decreasing $N_{\rm U}$, with minimum variance quantified by equation (\ref{eq:SFR_poisson_main}) or its more general version in Appendix \ref{sec:SFR_var_vs_NGMC}.

If we crudely assume that a galaxy must have at least $N_{\rm U}^{\rm crit} \approx 10$ Toomre-scale GBCs in order to exhibit time-steady star formation, then this implies that for a fiducial gas fraction $f_{\rm g}=0.2$ galaxies with circular velocity $v_{\rm c} \lesssim 46$ km/s should be bursty, regardless of redshift. 
For gas fractions of $f_{\rm g}\approx0.5$, which are common at high redshift, all galaxies with $v_{\rm c} \lesssim 116$ km/s should be bursty. 
Indeed, the observations compiled in \S \ref{sec:gas_fractions_app} of the Appendix show that gas fractions can be near unity at high redshift, in which case the star-forming discreteness effect can induce burtiness up to circular velocities $v_{\rm c}\sim 230$ km/s corresponding to massive galaxies. 
We stress, however, that our model also predicts a smooth transition from time-steady to bursty star formation with decreasing $v_{\rm c}$, which is consistent with the trends found in observations \citep[e.g.,][]{2012ApJ...744...44W, 2014MNRAS.441.2717K}. 

We can also express our predictions for $N_{\rm U}$ in terms of stellar mass using the model for the $M_{\star}-v_{\rm c}$ relation summarized in the previous section. 
This is done in Figure \ref{fig:N_GMC_vs_Mstar}. 
The curves in this figure depend on redshift because the conversion between $M_{\star}$ and $v_{\rm c}$ depends on redshift. 
To connect more directly with observations, the solid grey curves in each panel show our model predictions for the typical gas fractions as a function galaxy mass and redshift evaluated using the scaling relations summarized in Appendix \ref{sec:galaxy_scalings}.

\subsection{Feedback timescale and discreteness effects acting together: the effects of gas fractions}
\label{sec:gas_fraction}
One effect apparent in Figure \ref{fig:N_GMC_vs_Mstar} is that high-redshift galaxies (here with emphasis on $z=2$) can be bursty owing to both because their discs collapsing too rapidly for feedback to respond effectively \emph{and} because they contain few Toomre-scale GBCs at any given time. 
As mentioned above, this is because high-redshift galaxies have high typical gas fractions, $f_{\rm g}\gtrsim0.5$ being common \citep[e.g.,][]{2010ApJ...713..686D, 2010Natur.463..781T, 2013ApJ...768...74T}. 
This result is in agreement with rest-UV observations of massive high-redshift galaxies, which show a relatively small number of very massive ``giant'' star-forming clumps \citep[the analogs of local GMCs; e.g.,][]{2007ApJ...658..763E, 2011ApJ...733..101G, 2012ApJ...753..114W}. 
As a consistency check, present-day galaxies like the Milky Way should exhibit time steady star formation according to our model, and this is in agreement with high-resolution cosmological simulations that predict that such galaxies settle into a more steady mode of star formation at $z\lesssim 1$ \citep[][]{2014MNRAS.445..581H, 2015MNRAS.454.2691M, 2017MNRAS.466...88S}. 
\begin{figure}
\begin{center}
\mbox{
\includegraphics[width=0.475\textwidth]{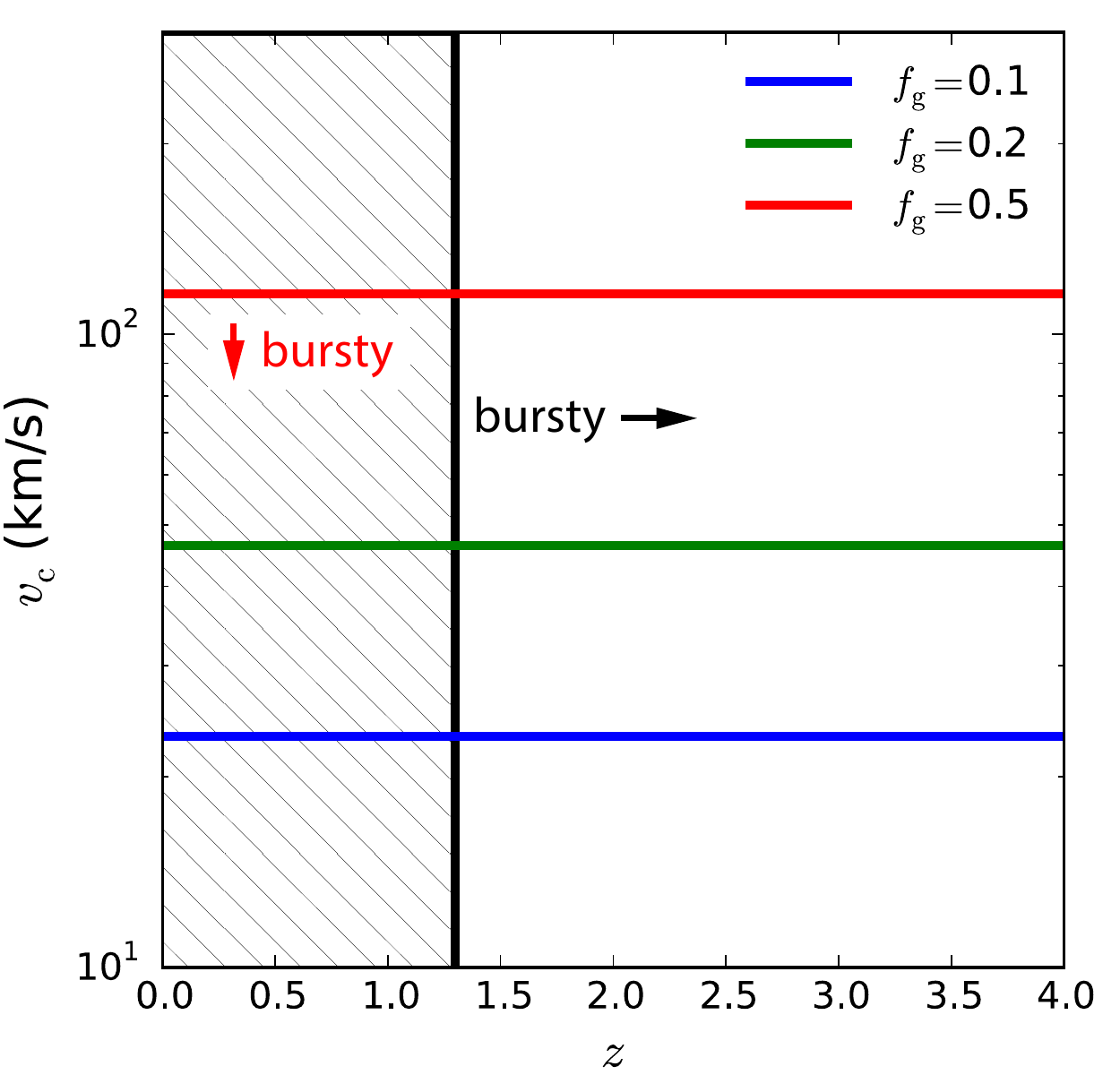}
}
\end{center}
\caption[]{Summary of where bursty star formation is expected in the space of galaxy circular velocity vs. redshift. 
All galaxies at $z\gtrsim1$ are expected to be bursty (the model predicts a transition at $z=1.3$). Below $z\sim1$, low-circular velocity galaxies are also bursty (depending on gas fraction). 
In this figure, the horizontal lines for different gas fractions correspond to $N_{\rm U} = 10$, but in reality a continuum of burstiness is expected, increasing with decreasing $N_{\rm U}$. 
The only galaxies in which stellar feedback can sustain steady star formation are relatively massive galaxies at $z\lesssim1$. 
}
\label{fig:bursty_space} 
\end{figure}

Figure \ref{fig:bursty_space} summarizes the parameter space in which we expect burstiness either due to short galactic dynamical timescales 
or to the small number of Toomre-scale GBCs. 
As the figure shows, our model predicts that only relatively massive galaxies at low redshift can sustain time-steady star formation.

In a different analytic analysis also expanding on the FG13 KS relation model, \cite{2017MNRAS.465.1682H} noted a different effect of gas fractions. 
At high gas fractions -- corresponding to thick discs with high turbulent velocities -- stellar feedback is predicted to be more efficient at driving galactic winds. 
\cite{2017MNRAS.465.1682H} argue that this effect explains why strong galactic winds are prevalent at high redshift but become weaker at late times in massive galaxies. 
When strong galactic winds are present, the fallback onto galaxies of outflows as they recycle \citep[e.g.,][]{2010MNRAS.406.2325O, 2017MNRAS.470.4698A} can enhance the time variability of star formation by sustaining a recurrent series of ``inflow-star formation-outflow'' cycles, on top of the gas accretion (and galaxy mergers) expected from the development of large-scale structure.

\section{Discussion and conclusions}
\label{sec:discussion}
By examining two limits in which a feedback-regulated model for the origin of the KS relation \citep[][]{2013MNRAS.433.1970F} 
breaks down, we have shown that the SFRs of galaxies are expected to be bursty at $z \gtrsim 1$ (for galaxies of all masses) and in dwarf galaxies (at all redshifts). 
At $z\gtrsim1$, the characteristic free fall time in galactic discs is generically shorter than the timescale $t_{\rm SN}\approx20$ Myr necessary for supernova feedback to output most of its energy. 
Therefore, SN feedback cannot establish a stable balance with gravity (\S \ref{sec:stellar_evolution}). 
Below $z \approx 1$, galaxies in which star formation is confined to just a small number Toomre-scale GBCs will inherit a time-dependent SFR from their GBCs. 
Only in galaxies in which a reasonably large number of Toomre-scale GBCs (quantified by the parameter $N_{\rm U}$) contribute to the galactic SFR can the total SFR become time steady by averaging (\S \ref{sec:discreteness}). 
Our model predicts that $N_{\rm U}$ decreases with galaxy mass (or, equivalently, galaxy circular velocity) and can thus explain why dwarf galaxies remain bursty all the way to $z=0$. 
The effect is enhanced by the high gas fractions in dwarf galaxies, which make individual Toomre-scale GBCs relatively massive compared to the stellar mass of the galaxy.

At high redshift, including around the peak of the cosmic star formation history at $z\sim2$ that is the focus of a number of large-scale observational efforts \citep[e.g.,][]{2011ApJS..197...35G, 2014ApJ...795..165S, 2015ApJS..218...15K, 2015ApJ...799..209W}, both the timescale and discreteness effects that we have identified should simultaneously occur owing to elevated gas fractions. 
In other words, high-redshift galaxies are expected to exhibit bursty star formation both because stellar feedback cannot respond sufficiently rapidly to gravitational collapse of the disc \emph{and} because star formation occurs in a small number of massive clumps. 

Overall, our model predicts that galaxies with time-steady SFRs are the exception rather than the norm, with only fairly massive galaxies at $z\lesssim 1$ (such as the Milky Way) being capable of sustaining time-steady star formation.  
We stress that on cosmological timescales, galactic SFRs are sustained by gas accretion from the intergalactic medium \citep[e.g.,][]{2005MNRAS.363....2K, 2009Natur.457..451D, 2011MNRAS.417.2982F, 2012MNRAS.421...98D}.
As a result, galaxy-averaged SFRs can evolve smoothly on cosmological timescales even in the regimes in which our model predicts variability on shorter timescales (the short timescale variability must average out to satisfy constraints on the evolution of the SFR on cosmological timescales imposed by the external gas supply). 
The amplitude of the SFR variability is increased by gas blowouts that follow star formation bursts (which can completely suppress star formation for a time) and gas fallback (which can fuel enhanced star formation). 
We also note that the causes of SFR variability identified in this paper are in addition to other known sources of star formation burstiness, such as galaxy mergers 
\citep[e.g.,][]{1991ApJ...370L..65B, 2010MNRAS.402.1693H, 2016MNRAS.462.2418S}. 

The mechanisms identified in this paper can induce star formation variability on a broad range of timescales. 
The shorter timescale variability should be apparent in indicators sensitive to star formation on $\lesssim5$ Myr timescales, such as H$\alpha$, but would be missed when SFRs are measured using indicators sensitive only to longer timescales, such as the UV continuum or its dust-processed counterparts in the infrared or millimeter ranges \citep[e.g.,][]{HaywardLanz2014}.
As shown in previous studies, the intrinsic SFR burstiness in nearby dwarf galaxies owing to the discreteness of star formation units (sometimes phrased in terms of stochastic sampling of star clusters) likely plays an important role in explaining the observed distribution of H$\alpha$-to-UV ratios \citep{2011ApJ...741L..26F, 2012ApJ...744...44W}, in addition to stochastic sampling of the IMF. 
Our model predicts that galaxies, including more massive ones, are also bursty at high redshift.  The intrinsic SFR variability should thus contribute to differences in SFRs inferred with different indicators at high redshift as well. 
One intriguing possibility is that recurrent star formation bursts in ordinary galaxies could explain substantial populations of extreme emission line galaxies observed at high redshift \citep[e.g.,][]{2011ApJ...742..111V, 2017ApJ...838L..12F}.

A testable prediction, for which there is already some support, is that the scatter in the SFR-$M_{\star}$ relation should be larger when measured using H$\alpha$ than with longer-timescale SFR indicators at high redshift and in dwarfs \citep[see][]{2017MNRAS.466...88S}. 
The transitions from bursty to time steady at late time and in massive galaxies predicted by the model are also found in recent cosmological hydrodynamic simulations that model stellar feedback in a spatially and time resolved manner \citep[e.g.,][]{2014MNRAS.445..581H, 2015ApJ...804...18A}. 
It is in fact a generic prediction of such simulations that the SFRs of relatively massive galaxies like the Milky Way only become time steady at $z \lesssim 1$. 
Lower-resolution simulations that do not explicitly resolve star formation into GBC units or do not model the time correct dependence of stellar feedback processes cannot capture the effects identified in this work. 
This is typically the case even in the latest, state-of-the-art large-volume cosmological simulations, which still have relatively coarse resolution in the ISM \cite[e.g.,][]{2014MNRAS.444.1518V, 2015MNRAS.446..521S, 2016MNRAS.462.3265D}. 
Because of this, current large-volume simulations likely predict star formation histories that are artificially smooth.

In this paper, we assumed that FG13's KS relation model could be applied to all galaxies, but that model was developed for high-$\Sigma_{\rm g}$ galaxies and assumes that the ISM is primarily supported by SN-driven turbulence. 
Our burstiness criteria could be quantitatively refined by considering modifications of the KS relation model involving other feedback processes and interstellar chemistry, which can be more important at lower $\Sigma_{\rm g}$.  
We nevertheless expect the arguments presented herein to hold approximately outside the range of strict applicability of the FG13 model. 
This is because a KS relation is \emph{observed} across essentially the entire galaxy population \citep[e.g.,][]{1998ApJ...498..541K, 2008AJ....136.2846B, 2010MNRAS.407.2091G} 
and previous theoretical studies have shown that the basic principles of star formation regulation by stellar feedback can be generalized to other feedback mechanisms, including stellar radiation \citep[][]{2005ApJ...630..167T, 2010ApJ...721..975O, 2017MNRAS.465.1682H}. 
Extensions of our feedback-regulated model that include additional processes such as radiation and ISM chemistry should be subject to similar failures of stable regulation as the ones identified for SN feedback in this paper, but may differ in their detailed predictions. 

Finally, we note that observations suggest that galactic winds are most prevalent in the high-redshift and dwarf regimes in which our model predicts bursty star formation \citep[e.g.,][]{2005ApJ...621..227M, 2010ApJ...717..289S, 2014ApJ...796..136B} and that galaxy-scale outflows become significantly weaker at late times in large galaxies \citep[e.g.,][]{2002ASPC..254..292H, 2015ApJ...809..147H}. 
Similar trends are found in the wind mass loading factors predicted by recent cosmological simulations that resolve the generation of galaxy-scale outflows from the injection of feedback energy on the scale of star-forming regions \citep[e.g.,][]{2015MNRAS.454.2691M, 2017MNRAS.470.4698A}. 
It would thus be very interesting to understand better the connection between SFR variability and galactic winds. 
One intriguing possibility is that powerful galactic winds are associated with galaxies in which SNe are strongly clustered in space and/or time (conditions associated with bursty star formation in our model). 
SN clustering can promote the formation of hot superbubbles that can more effectively break out of galactic discs than isolated supernova remnants and this may be key to driving powerful outflows \citep[e.g.,][]{2015MNRAS.450..504M, 2016MNRAS.456.3432G, 2016arXiv161008971L, 2017arXiv170401579F}. 

\section*{Acknowledgments}
We are grateful for useful discussions with Paul Torrey, Philip Hopkins, Eliot Quataert, Norman Murray, Christopher Hayward, Martin Sparre, and Alexander Gurvich. 
Alexander Richings, Christopher Hayward, and Zachary Hafen provided many useful comments on the manuscript. 
This work was supported by NSF through grants AST-1412836, AST-1517491, AST-1715216, and CAREER award AST-1652522, and by NASA through grant NNX15AB22G. 

\appendix

\section{Galaxy scaling relations}
\label{sec:galaxy_scalings}
To predict in which galaxies star formation is expected to be bursty, we need to know how galaxy properties evolve with mass and redshift. 
This appendix summarizes galaxy scaling relations that we use for this purpose. 

\begin{figure}
\begin{center}
\mbox{
\includegraphics[width=0.47\textwidth]{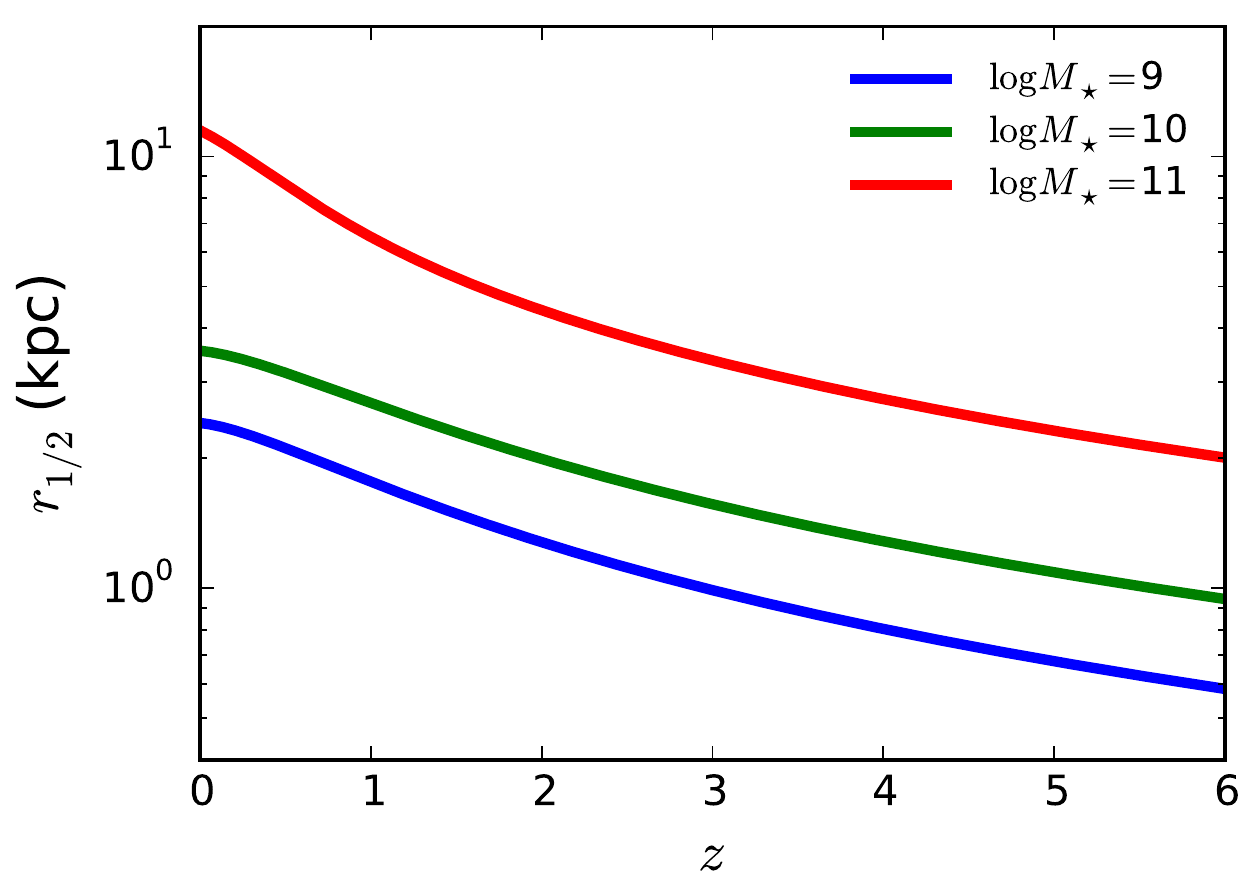}
}
\end{center}
\caption[]{Model half-mass radius of galaxy discs $r_{1/2}\approx0.02R_{\rm vir}$ as a function of redshift, for different fixed stellar masses. 
At fixed mass, galaxy discs are smaller at high redshift.
}
\label{fig:Rdisk_Shibuya} 
\end{figure}

\begin{figure*}
\begin{center}
\mbox{
\includegraphics[width=0.495\textwidth]{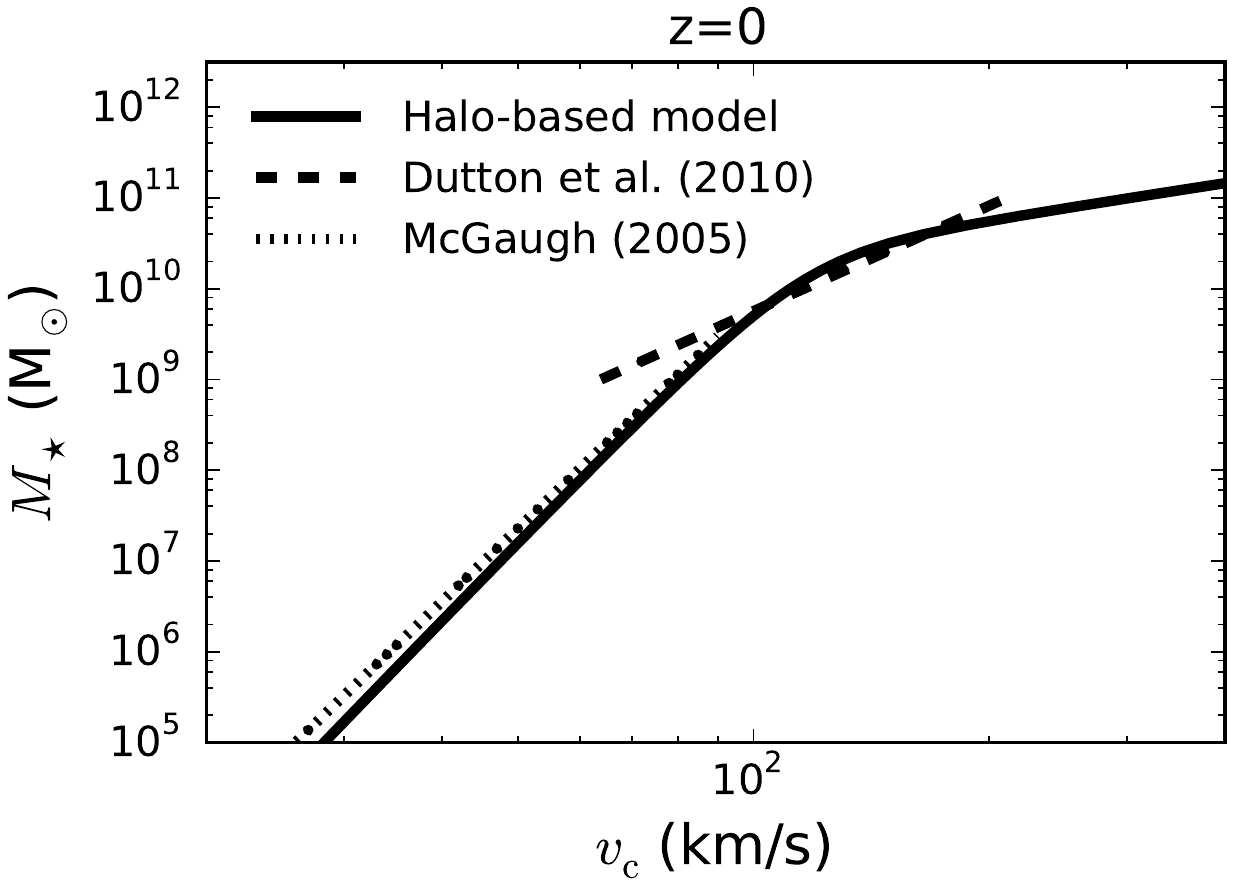}
\includegraphics[width=0.495\textwidth]{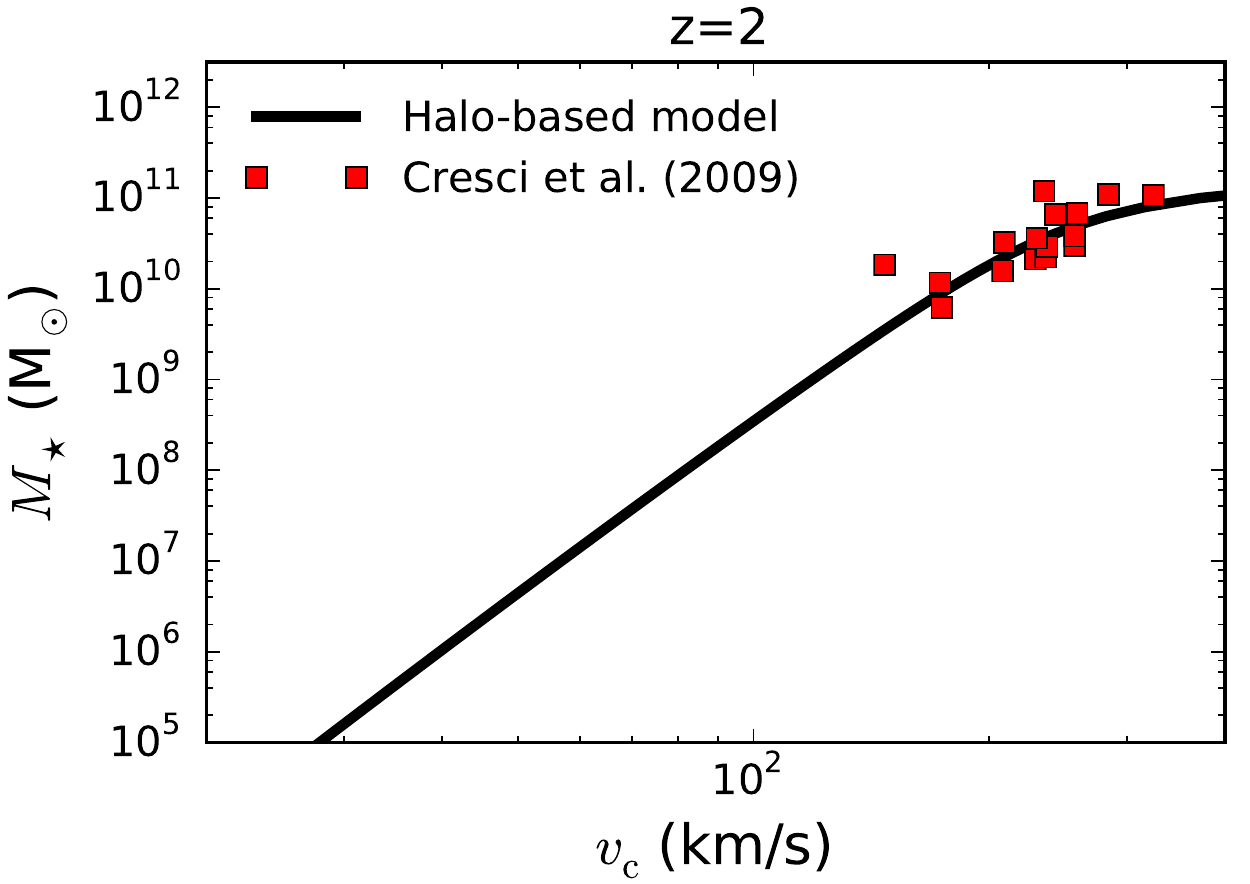}
}
\end{center}
\caption[]{Stellar mass vs. galaxy circular velocity. 
The solid curves show the halo-based model used in this paper. 
At $z=0$ (left), the model is compared to observational constraints on the Tully-Fisher relations from \cite{2005ApJ...632..859M} and \cite{2010MNRAS.407....2D}. 
At $z=2$, the model is compared to observations of massive galaxies at $1.5 \lesssim z \lesssim 2.5$ from \cite{2009ApJ...697..115C}. 
The model accurately reproduces the observed $z=0$ Tully-Fisher relation over a large dynamic range of galaxy masses and correctly predicts its evolution to $z=2$ (at least at the massive end).
}
\label{fig:TF_comparison} 
\end{figure*}

\subsection{Disc radii}
\label{sec:disc_radii}
For the disc radii, we follow the observational study of \cite{2015ApJS..219...15S} and assume that the half-mass radius of a galaxy\footnote{In \cite{2015ApJS..219...15S}, this result is demonstrated empirically for the effective radii of galaxies; here we identify the effective radius with the half-mass radius.} at any redshift is a constant fraction of the halo virial radius, $r_{1/2} \approx 0.02 R_{\rm vir}$.
We use the \cite{1998ApJ...495...80B} halo definition, in which the mean enclosed over-density depends on redshift: 
\begin{equation}
\Delta_{\rm c} = 18 \pi^{2} + 82d - 39d^{2},
\end{equation}
where, in a flat Universe with $\Omega_{\rm m} + \Omega_{\Lambda}=1$,
\begin{equation}
d \equiv \Omega_{\rm m}^{z} - 1
\end{equation}
and
\begin{equation}
\Omega_{\rm m}^{z} = \frac{\Omega_{\rm m} (1 + z)^{3}}{\Omega_{\rm m} (1+z)^{3} + \Omega_{\Lambda}}.
\end{equation}
The halo virial mass and radius are related by
\begin{equation}
M_{\rm vir} = \frac{4 \pi}{3} R_{\rm vir}^{3} \Delta_{\rm c}(z) \rho_{\rm c}(z),
\end{equation}
where
\begin{equation}
\rho_{\rm c} = \frac{3 H^{2}}{8 \pi G}
\end{equation}
is the critical density. 
At high redshift, $\Delta_{\rm c} \to 178$ but $\Delta_{\rm c}(z=0)\approx 97$. 
For any halo, we define the virial velocity as
\begin{equation}
V_{\rm vir} = \left( \frac{G M_{\rm vir}}{R _{\rm vir}} \right)^{1/2}.
\end{equation}

To connect to observations, we also require a conversion between halo mass and the stellar mass of the central galaxy. 
For this, we use the fitting formula as a function of both mass and redshift from the abundance matching analysis of \cite{2013MNRAS.428.3121M} for $M_{\star}-M_{\rm vir}$. 
Figure \ref{fig:Rdisk_Shibuya} shows the predicted evolution of galaxy discs sizes as a function of redshift, for different fixed stellar masses. 
Galaxies of a given mass are smaller at high redshift because $R_{\rm vir} \propto (1+z)^{-1}$ as $z\to \infty$ (at fixed halo mass). 

\subsection{Disc circular velocities}
\label{sec:TF}
To obtain more accurate results, it is useful to use the galaxy rotation velocity rather than the virial velocity of the parent halo in our calculations (although the two are similar to order unity). 
To do this, we derive a simple model calibrated to observations of the Tully-Fisher relation to capture the mass and redshift evolution of the relation between galaxy rotation velocity, $v_{\rm c}$, and stellar mass. 
Our model is based on the analysis of the evolution of halos in cold dark matter cosmologies by \cite{2001ApJ...555..240B}. 

For halos that follow an NFW profile \citep[][]{1997ApJ...490..493N}, the maximum circular velocity in the halo, $V_{\rm max}$, is related to the halo virial velocity through
\begin{equation}
\frac{V_{\rm max}^{2}}{V_{\rm vir}^{2}} \approx 0.216 \frac{c_{\rm vir}}{A(c_{\rm vir})},
\end{equation}
where $c_{\rm vir}$ is the halo concentration and 
\begin{equation}
\label{eq:A}
A(c_{\rm vir}) \equiv \ln{(1+c_{\rm vir})} - \frac{c_{\rm vir}}{1+c_{\rm vir}}.
\end{equation}
For the concentration, we use
\begin{equation}
\label{eq:concentration}
c_{\rm vir} \approx 15 \left( \frac{M_{\rm vir}}{\rm 10^{12}~M_{\odot}} \right)^{-0.2} (1+z)^{-1}.
\end{equation}
At $z=0$, this matches the mass dependence from \cite{2000MNRAS.318..203S} and at higher redshift this follows the scaling $c_{\rm vir} \propto (1+z)^{-1}$ proposed by \cite{2001ApJ...555..240B}. 
We assume that the galaxy circular velocity at $r_{1/2}$ ($v_{\rm c}$) is a constant fraction of $V_{\rm max}$,
\begin{equation}
v_{c} = \beta V_{\rm max},
\end{equation}
and calibrate $\beta$ to observations of the Tully-Fisher relation. 
As we show below, $\beta=0.9$ provide a good fit to both $z\approx0$ and $z\approx 2$ observational constraints, and so we adopt this value. 

With these ingredients, we can compute the galaxy circular velocity for any combination of stellar mass and redshift. 
In Figure \ref{fig:TF_comparison}, we plot $M_{\star}$ vs. $v_{\rm c}$ for our model at $z=0$ and $z=2$. 
At $z=0$, we also show the best-fit Tully-Fisher relation from \cite{2010MNRAS.407....2D} for massive galaxies and an approximate fit to lower-mass galaxies 
 to the observations compiled by \cite{2005ApJ...632..859M},
\begin{equation}
M_{\star}\approx 3\times10^{9}~{\rm M_{\odot}} \left( \frac{v_{\rm c}}{\rm 90~km/s} \right)^{8.3}
~~~~(v_{\rm c}<{\rm~90 km/s}).
\end{equation}
This fit to the \cite{2005ApJ...632..859M} data is only approximate because many of the low-mass galaxies in the sample only have an upper limit on their stellar mass.
At $z=2$, we show stellar masses and circular velocities measured using integral field observations of massive galaxies from the SINS survey by \cite{2009ApJ...697..115C}. 
As the figure shows, our simple model for the $M_{\star}-v_{\rm c}$ relation provides a good fit to both the $z=0$ and $z=2$ observations. 

\subsection{Gas fractions}
\label{sec:gas_fractions_app}
\begin{figure}
\begin{center}
\mbox{
\includegraphics[width=0.47\textwidth]{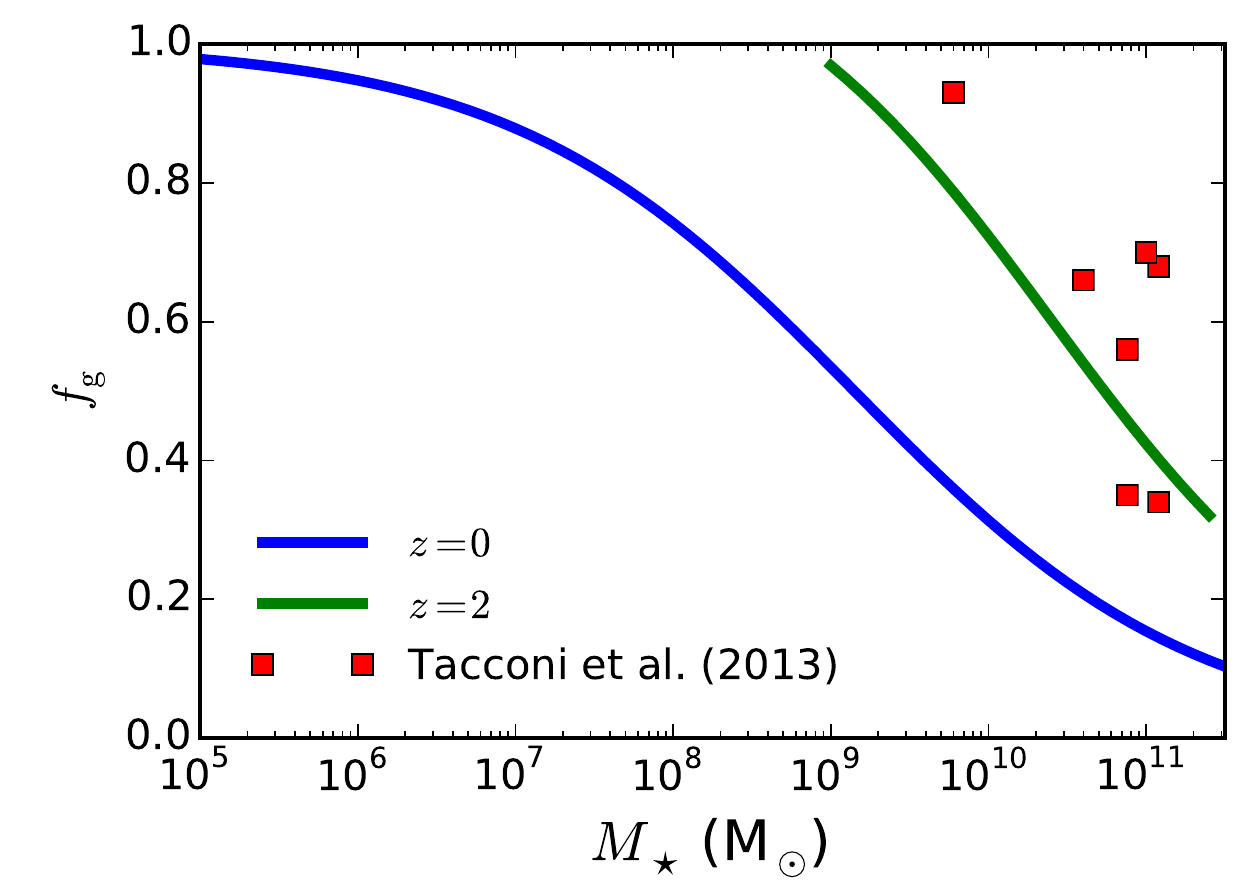}
}
\end{center}
\caption[]{Gas mass fraction as a function of galaxy stellar mass used to evaluate our model predictions at $z=0$ and $z=2$ (solid curves). 
The red squares show gas fractions derived from observations of disc galaxies in \cite{2013ApJ...768...74T} in the redshift interval $1.5<z<2.5$.
}
\label{fig:fg_vs_Mstar} 
\end{figure}

For the gas fractions, we start with the fitting formulae from \cite{2009ApJ...691.1424H}: 
\begin{align}
f_{\rm g,0} = \left[1 + \left( \frac{M_{\star}}{\rm 10^{9.15}~M_{\odot}} \right)^{0.4} \right]^{-1}\\ \notag
f_{\rm g}(M_{\star},~z) = f_{\rm g,0} \left[ 1 - \tau(z) (1 - f_{0}^{3/2}) \right]^{-2/3},
\end{align}
where $\tau(z)$ is the fractional lookback time to redshift $z$ (such that $\tau(z=0)=0$ and $\tau(z\to \infty) = 1$). 
At $z=2$, we find that multiplying the $f_{\rm g}$ implied by these expressions by 1.2 provides a better fit to the more recent gas fraction measurements compiled in \cite{2013ApJ...768...74T}.
We therefore multiply $f_{\rm g}$ by this factor for our $z=2$ predictions (but not for $z=0$). 
At this redshift, we only consider relatively massive galaxies for which there are good gas fraction measurements ($M_{\star} \sim 10^{9}-10^{11}$ M$_{\odot}$). 
Figure \ref{fig:fg_vs_Mstar} shows the gas fractions used to evaluate our model predictions as a function of stellar mass, at $z=0$ and $z=2$. 
In this figure, the data points from \cite{2013ApJ...768...74T} are restricted to galaxies in the redshift interval $1.5<z<2.5$ and that are classified as discs (we exclude galaxies classified as mergers or dispersion dominated). 
Two trends are noteworthy: gas fractions increase with decreasing stellar mass and with increasing redshift. 
These gas fraction trends drive the burstiness due to star formation discreteness in dwarf galaxies and in high-redshift galaxies predicted in \S \ref{sec:discreteness}.

\section{SFR variance vs. $N_{\rm U}$}
\label{sec:SFR_var_vs_NGMC}
In \S \ref{sec:discreteness}, we related the burstiness of star formation to the expected number of most massive (Toomre-scale) GBCs, $N_{\rm U}$, in a galaxy. 
In this section, we show that the fractional variance of the instantaneous galaxy SFR is primarily determined by $N_{\rm U}$ under general assumptions consistent with the star formation model on which this paper is based (see \S \ref{sec:preliminaries}). 
Specifically, we derive an explicit expression for 
\begin{equation}
\frac{\sigma_{\rm SFR}}{\langle SFR \rangle}
\end{equation}
in terms of $N_{\rm U}$, where $\langle SFR \rangle$ is the mean SFR in the galaxy (averaged over a long timescale $\gg100$ Myr) and $\sigma_{\rm SFR}$ is the standard deviation of the instantaneous SFR. 
 
We start with a general function describing the time evolution of the SFR formation rate within an individual GBC as a function of mass $m$ and time $t$ since GBC formation, $SFR_{i}(m,~t)$. 
We will later evaluate our final result for different variations of $SFR_{i}$. 
The GBC mass function 
\begin{equation}
\frac{dN}{dm} = \left\{
\begin{array}{cc}
A m^{-\alpha} & ~M_{\rm L} \leq m \leq M_{\rm U}\\
0 & \text{otherwise,}
\end{array} \right.
\end{equation}
describes the distribution of GBCs present in the galaxy at any given time. 
At any time, each such GMC will be ``caught'' at a random time $0 \leq t < t_{\rm GBC}$ during its evolution, where $t_{\rm GBC}$ is the GBC lifetime. 
The total galactic SFR is, by definition, the sum over the star formation rates of GBCs in the galaxy:
\begin{equation}
SFR = \Sigma_{i=1}^{N_{\rm tot}} SFR_{i}.
\end{equation}
We treat $SFR_{i}$ as a function of two random variables, $m$ and $t$. 
For each $i$, $m$ and $t$ are drawn from a joint distribution
\begin{equation}
f_{m,t}(m,~t)=f_{m}(m)f_{t}(t)
\end{equation}
which factors into marginal probability density functions for $m$ and $t$ since these are independent random variables. 

The probability density function for $m$ is proportional to the GMC mass function, but normalized such that $\int dm f_{m}(m)$=1:
\begin{align}
f_{m}(m) = \left\{
\begin{array}{cc}
\tilde{A} m^{-\alpha} & ~M_{\rm L} < m \leq M_{\rm U}\\
0 & \text{otherwise,}
\end{array} \right.
\end{align}
where
\begin{align}
\tilde{A} &= \frac{(\alpha - 1) (M_{\rm L} M_{\rm U})^{\alpha-1}}{M_{\rm U}^{\alpha-1}-M_{\rm L}^{\alpha-1}} \\
&= \frac{\alpha-1}{(\beta^{1-\alpha}-1)} M_{\rm  U}^{\alpha-1}.
\end{align}
In the last expression, we defined $\beta \equiv M_{\rm L}/M_{\rm U}$, which will simplify later results. 
The probability density function for $t$ is simply uniform over the GBC lifetime:
\begin{equation}
f_{t}(t) = \left\{
\begin{array}{cc}
1/t_{\rm GBC} & 0 \leq t \leq t_{\rm GBC}\\
0 & \text{otherwise.}
\end{array}
\right.
\end{equation}

We are now ready to evaluate $\langle SFR \rangle$ and $\sigma_{\rm SFR}$:
\begin{align}
\label{eq:mean_SFR1}
\langle SFR \rangle = N_{\rm tot} \langle SFR_{i} \rangle,
\end{align}
where
\begin{align}
N_{\rm tot} & = \int_{M_{\rm L}}^{M_{\rm U}} dm \frac{dN}{dm} \\
\label{eq:Ntot_eval}
&= 
A\frac{(\beta^{1-\alpha}-1)}{\alpha-1} M_{\rm U}^{1-\alpha}.
\end{align}
For the variance, we start with the general result
\begin{align}
\label{eq:variance_gen}
\sigma_{\rm SFR}^{2} = \langle SFR^{2} \rangle - \langle SFR \rangle^{2},
\end{align}
and proceed to evaluate $\langle SFR^{2} \rangle$. We have
\begin{align}
SFR^{2} = \Sigma_{i}^{N_{\rm tot}} SFR_{i}^{2} + \Sigma_{i\neq j} SFR_{i} SFR_{j}
\end{align}
and therefore, since $SFR_{i}$ and $SFR_{j}$ are independent but identically distributed for $i \neq j$,
\begin{align}
\label{eq:mean_SFR2}
\langle SFR^{2} \rangle = N_{\rm tot} \langle SFR_{i}^{2} \rangle + N_{\rm tot}(N_{\rm tot}-1) \langle SFR_{i} \rangle^{2}.
\end{align}
As before, the problem reduces to evaluating moments of $SFR_{i}$, the SFR evolution of an individual GBC.

Using equations (\ref{eq:mean_SFR1}), (\ref{eq:variance_gen}), and (\ref{eq:mean_SFR2}), we find that
\begin{align}
\frac{\sigma_{SFR}^{2}}{\langle SFR^{2} \rangle} &= \frac{1}{N_{\rm tot}} \frac{\langle SFR_{i}^{2} \rangle}{\langle SFR_{i}\rangle ^{2}} + \frac{N_{\rm tot}(N_{\rm tot}-1)}{N_{\rm tot}^{2}} - 1\\
\label{eq:variance_convenient}
& \approx \frac{1}{N_{\rm tot}} \frac{\langle SFR_{i}^{2} \rangle}{\langle SFR_{i}\rangle ^{2}}.
\end{align}
The last approximation holds so long as $N_{\rm tot} \gg 1$. 
To connect with the results of \S \ref{sec:discreteness}, we must express our results in terms of $N_{\rm U}$ rather than $N_{\rm tot}$. 
By definition,
\begin{align}
N_{\rm U} &= \frac{\int_{M_{\rm L}}^{M_{\rm U}}dm (dN/dm) m}{M_{\rm U}} \\
& = 
A \frac{(1-\beta^{2-\alpha})}{2-\alpha} M_{\rm U}^{1-\alpha},
\end{align}
and therefore, using equation (\ref{eq:Ntot_eval}),
\begin{align}
N_{\rm tot} &= 
\frac{(2-\alpha)}{(\alpha-1)} \frac{(\beta^{1-\alpha}-1)}{(1 - \beta^{2-\alpha})} N_{\rm U}.
\end{align}
Equation (\ref{eq:variance_convenient}) is convenient because it allows us to calculate how the fractional variance of the galactic SFR depends on parameters of the GBC mass function and on the function $SFR_{i}$ describing the time evolution of individual GBCs.

Consider a specific, but rather general parameterization of GBC evolution:
\begin{equation}
\label{eq:SFR_i}
SFR_{i}(m,~t) = \left\{
\begin{array}{cc}
0 & -(1-\gamma)t_{\rm GBC} \leq t < 0\\
B m t^{\delta} & 0 \leq t \leq \gamma t_{\rm GBC}.
\end{array}
\right.
\end{equation}
In this expression, $t_{\rm GBC}$ is the total GBC lifetime. 
The parameter $\gamma$ defines the fraction of $t_{\rm GBC}$ during which the GBC forms stars at a significant rate. 
This parameter allows us to model a possible early phase during which, for example, strong turbulence generated during the assembly of the GBC prevents star formation from occurring by maintaining a high virial parameter \citep[e.g.,][]{2012ApJ...759L..27P}. 
The proportionality to $m$ captures the fact that, on average, we expect the SFR of an individual GBC to scale with its mass. This is true, for example, if $SFR_{i} \sim m/t_{\rm GBC}$. 
Finally, the power-law index $\delta$ characterizes the SFR evolution as the cloud collapses.  
Recent simulations suggest that to a good approximation the SFR within GBCs increases linearly with time ($\delta \approx 1$) after the first stars form \citep[][]{2015ApJ...800...49L, 2016ApJ...829..130R, 2016arXiv161205635G}.  
Real GBC are likely disrupted more gradually than the abrupt cut off at the end implied by the simple model in equation (\ref{eq:SFR_i}). 
For simplicity, we do not model this explicitly here but note that the sensitivity of our main result below on different $SFR_{i}$ parameterizations can be gauged from the dependence on the parameters $\gamma$ and $\delta$. 
The constant prefactor $B$ defines the absolute normalization of the SFR; its value has no impact on the following result as it cancels out exactly in equation (\ref{eq:variance_gen}). 

For the $SFR_{i}$ in equation (\ref{eq:SFR_i}), we find
\begin{align}
\frac{\sigma_{\rm SFR}^{2}}{\langle SFR \rangle^{2}} = 
\frac{1}{\gamma N_{\rm U}}
\frac{(1+\delta)^{2}}{(1+2\delta)}
\left( \frac{2-\alpha}{3-\alpha} \right) 
\left(
\frac{1-\beta^{3-\alpha}}{1-\beta^{2-\alpha}}
\right).
\end{align}
This expression involves several parameters but its numerical value is not very sensitive to most. 
For example,
\begin{align}
\frac{\sigma_{\rm SFR}}{\langle SFR \rangle} = \frac{S}{\sqrt{\gamma N_{\rm U}/0.5}},
\end{align}
where $S \approx 1.1,~1.1,~1.0,~0.9,~0.8$ and $1.2$ for $(\alpha,~\beta,~\delta)=(1.5,~0.1,~1)$, $(1.8,~0.1,~1),$ $(2,~0.1,~1),$ $(1.8,~0.01,~1),$ $(1.8,~0.001,~1)$ and $(1.8,~0.1,~2)$.

This result has a simple interpretation. 
In our model, the total SFR is dominated by Toomre-scale GBCs. 
Thus, at any given time the galactic SFR is roughly a sum over $\sim \gamma N_{\rm U}$ Toomre-scale GBCs, each caught at a random time during its evolution after star formation has begun. 
Since in individual GBCs the SFR is assumed to increase with time (before disruption by feedback), the total SFR can be approximated as a sum over some GBCs that are effectively in a ``low SFR'' state and others that are in a ``high SFR'' state. 
The number of ``high SFR'' GBCs varies with time due to stochastic effects, with a Poisson fractional standard deviation $\propto 1 / \sqrt{\gamma N_{\rm U}}$. 
This quantitatively illustrates our claim in the main text that the variability of the galactic SFR increases with decreasing $N_{\rm U}$. 

\bibliography{references}

\end{document}